\documentclass[twoside,12pt]{article}
\usepackage{amsmath}
\usepackage{amssymb}
\usepackage{latexsym}
\usepackage{epsfig}
\usepackage{cite}
\usepackage{graphics}
\newlength{\dinwidth}
\newlength{\dinmargin}
\newtheorem{theorem}{Theorem}[section]
\newtheorem{prop}[theorem]{Proposition}
\newtheorem{lemma}[theorem]{Lemma}
\newtheorem{cor}[theorem]{Corollary}

\newenvironment{proof}{\medskip \noindent 
            {\bf Proof.}}{ \hfill $\square$ \medskip}

\def\wnet{{\{\As(W)\}_{W \in\Ws}}}

\def\msc{{Modular Stability Condition}}
\def\diag{{\textnormal{diag}}}

\def\Im{{\textnormal{Im}}}
\def\ad{{\textnormal{ad}}}
\def\As{{\cal A}}
\def\Bs{{\cal B}}

\def\Es{{\cal E}}

\def\Hs{{\cal H}}

\def\Ms{{\cal M}}
\def\Ns{{\cal N}}
\def\Os{{\cal O}}

\def\Rs{{\cal R}}

\def\Us{{\cal U}}

\def\Ws{{\cal W}}

\def\RR{{\mathbb R}}
\def\IN{{\mathbb N}}

\def\CC{{\mathbb C}}

\def\wnet{{\{\As(W)\}_{W \in\Ws}}}

\def\diag{{\textnormal{diag}}}

\def\sigmavec{{\boldsymbol{\sigma}}}
\newcommand{\AdS}{\mbox{{AdS}$^n\,$}}  
\newcommand{\AdSt}{\mbox{{AdS}$^2\,$}}  
\newcommand{\AdSG}{\mbox{SO$_0(2,n\!-\!1)$}}
\newcommand{\AdSGfour}{\mbox{SO$_0(2,3)$}}
\newcommand{\AdSGG}{\mbox{SO$(2,n\!-\!1)$}}
\newcommand{\CA}{{\cal A}}
\newcommand{\CH}{{\cal H}}

\newcommand{\rva}{{\Omega \rangle}}
\newcommand{\rvap}{{\Omega_p \rangle}}
\newcommand{\lva}{{\langle\Omega }}
\newcommand{\lvap}{{\langle\Omega_p }}

\newcommand{\ie}{{\it i.e. \,}}

\begin{document}
\title{Stable Quantum Systems in Anti--de Sitter Space: Causality, 
Independence and Spectral Properties} 

\author{{\Large Detlev Buchholz\,$^a$ \ and \  
Stephen J.\ Summers\,$^b$ }\\[5mm]
${}^a$ Institut f\"ur Theoretische Physik, 
Universit\"at G\"ottingen, \\ 37077 G\"ottingen, Germany  \\[2mm]
${}^b$ Department of Mathematics, 
University of Florida, \\ Gainesville FL 32611, USA}

\date{\small Dedicated to Jacques Bros on the occasion of his
  seventieth birthday} 

\maketitle 

\abstract{\noindent If a state is passive for uniformly accelerated
observers in $n$-dimensional ($n \geq 2$) anti--de Sitter space--time
({\em i.e.} cannot be used by them to operate a {\em perpetuum
mobile}), they will (a) register a universal value of the
Unruh temperature, (b) discover a PCT symmetry, and (c) find
that observables in complementary wedge--shaped regions necessarily
commute with each other in this state. The stability properties of such a
passive state induce a ``geodesic causal structure'' on AdS and
concommitant locality relations. It is shown that observables in
these complementary wedge--shaped regions fulfill strong additional
independence conditions. In two-dimensional AdS these even suffice to
enable the derivation of a nontrivial, local, covariant net indexed by
bounded spacetime regions. All these results are model--independent 
and hold in any theory which is compatible with a weak notion of 
space--time localization. Examples are provided of models satisfying 
the hypotheses of these theorems.}

\section{Introduction and basic assumptions} \label{intro}
\setcounter{equation}{0}

     Quantum field theory in anti-de Sitter space--time (AdS) has been
studied for almost 40 years (see {\it e.g.} \cite{F1,AIS}), primarily
because it was found that AdS occurs as the ground state geometry in
certain supergravity theories with gauged internal symmetry
\cite{BrFr,dWNi2}. But it has become the object of an extraordinary
amount of attention since the AdS-CFT correspondence has 
emerged.\footnote{We refer the interested reader to the
SPIRES database, where a comprehensive list of articles on this topic can
be retrieved.} There is therefore motivation to clarify in a
model--independent setting and in a mathematically rigorous manner the
universal properties of such theories, as implied by generally
accepted and physically meaningful assumptions. This investigation has
lead us to results which apparently have not been remarked
in any form in the literature before.

     AdS is a maximally symmetric and globally static solution of the vacuum
Einstein equations. We consider here AdS of any dimension $n \geq 2$,
except when explicitly stated otherwise. 
It can conveniently be described in terms of Cartesian coordinates in the 
ambient space $\RR^{n+1}$ as the quadric surface
\begin{equation}
\AdS = \{x\in\RR^{n+1} \mid 
x^2 \doteq x_0^2 - x_1^2 - \dots - x_{n-1}^2 + x_n^2 = R^2 \} 
\end{equation}
with metric $g = \mbox{diag}(1,-1,\dots,-1,1)$ in diagonal form.
As the value of the radius $R$ is not relevant for the results of this 
paper, we shall set it equal to $1$ for convenience. The \AdS
isometry group is O$(2,n-1)$ whose identity component will be denoted
by \AdSG. \AdS is a homogeneous space of the group \AdSGG.  It is not
globally hyperbolic; indeed, it has closed timelike curves and has a
timelike boundary at spatial infinity through which physical data can
propagate. Although the covering space of \AdS eliminates the closed
timelike curves, it still has a timelike boundary at spatial
infinity. We shall find some notable differences between the
properties of quantum field theories on \AdS and those of theories on
the covering space.

    As some of our basic assumptions are motivated by physical
considerations concerning certain families of observables, we must 
collect some basic facts about observers in AdS. Let
$x_O \in \AdS$ be any point and let $\lambda (t)$, $t \in \RR$, be any
one--parameter subgroup of $\AdSG$ such that $t \mapsto \lambda(t) x_O$ 
is an orthochronous curve. (Note that AdS is time orientable.)
We interpret this curve as the worldline of some observer. 
Among these observers will be those moving along a geodesic (henceforth, 
geodesic observers) and those experiencing a constant acceleration (uniformly
accelerated observers). Points in a neighborhood of $x_O$ will, in
general, also give rise to orthochronous curves under the action of
the chosen subgroup of $\AdSG$, and we denote by $W$ the connected
neighborhood of $x_O$ in \AdS consisting of all such
curves. Typically, $W$ is the causal completion of the originally
specified worldline. We view the region $W$ as
the maximal possible localization for any laboratory within the
purview of the given observer. The associated dynamics are given by 
$e^{\, itM} \doteq U(\lambda(t))$ with suitable generator $M$. Since we are 
choosing a fixed parametrization of the pertinent subgroups of \AdSG, 
the proper time of the observer is obtained by rescaling $t$ with 
$((\dot{\lambda}(0) \, x_O)^2){}^{1/2}$.

     To become more precise, the geodesics of \AdS are conic sections 
by two--planes containing the origin of the ambient space $\RR^{n+1}$. 
So, one-parameter subgroups\footnote{See Appendix \ref{grouptheory} 
for our notation concerning $\AdSG$}  $\lambda(t)$, $t \in \RR$, 
of $\AdSG$ of the form $\lambda \lambda_{0n}(t) \lambda^{-1}$, 
$t \in \RR$, for some $\lambda \in \AdSG$ generate admissible 
geodesic worldlines in the sense just indicated. These worldlines are
closed, timelike curves, whose causal completion is the entire space
$\AdS$.  Hence, the maximal laboratory localization region $W$ for
such geodesic observers must be the entire space-time, $\AdS$. 
For uniformly accelerated observers, the corresponding 
one-parameter subgroups are of the form $\lambda \lambda_{01}(t) \lambda^{-1}$,
$t\in\RR$, for some $\lambda\in\AdSG$. Their laboratory regions,
called AdS wedges, are described immediately below. The algebras 
$\As(W)$ corresponding to any such wedge region as well as to
$W = \AdS$ are taken to be weakly closed. 

     We define a ``wedge'' in \AdS to be the causal completion of the
worldline of a uniformly accelerated observer in \AdS. To be concrete 
and in order to simplify the necessary computations, we consider the 
particular choice of region
\begin{equation}
W_R = \{ x \in \AdS \mid x_1 > |x_0|\, , x_n > 0 \} \, ,
\end{equation}
on which the one-parameter subgroup of boosts $\lambda_{01}(t)$, 
$t\in\RR$, in the $0$--$1$--plane acts in an orthochronous manner. For
any $x_O \in W_R$, the curve $t \mapsto \lambda_{01}(t) x_O$ is
the worldline of a uniformly accelerated observer for which the causal
completion is precisely $W_R$. By the assumed \AdSG \, covariance, 
all results concerning this wedge have natural extensions to all 
images of $W_R$ under \AdSG . We therefore define the set of \AdS 
wedges to be
\begin{equation}
\Ws \doteq \{ \lambda W_R \mid \lambda \in \AdSG \} \, . 
\end{equation}
These are maximal laboratory localizations for the uniformly
accelerated observers. 
 
     We can now describe the four standing assumptions of this paper. 
A discussion of their physical motivation is given in \cite{AdS},
so we shall only expand upon the less familiar ones. 

\begin{enumerate}
\item[(i)] There exists a strongly continuous, unitary, nontrivial 
representation $U$ of the symmetry group \AdSG \, acting on a separable
Hilbert space $\CH$.\footnote{It is sufficient here to 
consider the subspace of ``bosonic'' states, so we shall not 
need to proceed to the covering group of the spacetime 
symmetry group.}
\item[(ii)] On $\Hs$ act the global von Neumann algebra of observables 
$\CA = \As(\AdS)$, which contains any observable measurable in $\AdS$, 
and an isotonous family of von Neumann algebras $\wnet$ associated 
with the wedges $\Ws$. Furthermore, one has
\begin{equation} \label{weakadditivity}
\bigvee_{W \in \Ws} \As(W) = \As \, .
\end{equation}
\item[(iii)] For each wedge $W \in \Ws$ and $\lambda \in \AdSG$, 
one has the equality 
\begin{equation}
U(\lambda)  \CA (W) U(\lambda)^{-1} = \CA (\lambda W) \, .  
\end{equation}
\end{enumerate}

     The weak additivity condition (\ref{weakadditivity}) is a 
generalization of the natural idea that all observables are
constructed out of local ones. But in contrast to
\cite{AdS}, we do not assume that all observables can be constructed
out of observables with arbitrarily small localization region. This
is because there are nets of physical interest on curved space--times 
for which the condition (v) specified below holds but the algebras 
$\As(\Os)$ associated with all bounded open regions $\Os$ are trivial 
\cite{BMS,Rehren}. Also, there exist examples in which the algebra 
$\As(\Os)$ is nontrivial only for sufficiently large bounded regions 
$\Os$ \cite{BMS}. In both of these cases the assumption made in
\cite{AdS} is violated. We have therefore eliminated all assumptions 
referring to bounded regions. 
 
     We emphasize that we do not postulate from the outset any local
commutation relations of the observables. For, in contrast to the case
of globally hyperbolic space--times, the principle of Einstein
causality does not provide any clues as to which observables in AdS
should commute with each other. Instead, we shall {\it derive} such
commutation relations from stability properties of the vacuum, which
we now specify. 

     We shall assume that the state $\omega$ determined by $\Omega$ 
is passive (cf. \cite{PuWo} and Section 5.4.4 in \cite{BratRob}) 
for the dynamical system $(\As(W), \ad U(\lambda(t)))$, for all geodesic
and all uniformly accelerated observers described above. We recall 
that passivity is an expression of the Second Law of Thermodynamics. 
Since the vacuum is the most elementary system, all order parameters 
should have sharp values in this state. This is expressed by the 
weak mixing property:
\begin{equation} \label{weakmixing}
\lim_{T \rightarrow \infty}
\frac{\mbox{\footnotesize $1$}}{\mbox{\footnotesize $T$}}
\int_0^T  \, \left( \omega( A(t) B)  -  
\omega(A(t))\, \omega(B) \right) \, dt = 0 \, ,  
\end{equation}
for all $A,B \in \CA$, where  
$A(t) \doteq e^{\, itM } A e^{\, -itM }$. The restriction of the 
state $\omega$ to $\As(W)$ is said to be central if
$\omega(AB) = \omega(BA)$, for all $A,B \in \As(W)$. If this holds, 
then either $\Omega$ is
annihilated by most of the observables in $\As(W)$ or $\As(W)$ is
a finite algebra (cf. Section 8.1 in \cite{KadRing}). In quantum field 
theory this is a physically pathological circumstance, which we 
shall exclude from consideration. 

     These basic features of the vacuum can be summarized as follows.

\medskip

(iv) The vacuum vector $\Omega$ is cyclic for $\As$ and determines a 
passive, weakly mixing and noncentral state $\omega$ for all geodesic
and all uniformly accelerated observers.

\medskip

     The Standing Assumptions (i)--(iv) are model--independent
and physically natural. In this paper we shall show that these
assumptions entail that for geodesic observers the vacuum $\omega$
is a ground state; uniformly accelerated observers in AdS will
register a universal value of the Unruh temperature; they
will discover a PCT symmetry; and they will find that observables
localized in complementary wedge--shaped regions must commute in the
vacuum state.  Not only do such observables commute in this sense, 
but the corresponding algebras manifest strong properties of 
statistical independence, the nature of which will be studied in 
detail.  We shall also see that these assumptions imply that quantum 
theories on AdS obey a geodesic causal structure. 

     Related results appeared in \cite{AdS}, and we revisit some of
those arguments here in more detail than in that announcement. But our
research in the intervening time has led not only to further results
and a weakening of the assumptions, but also to a shift in our point of
view, which now places emphasis on the locality and independence
properties which can be derived from our assumptions. We establish
independence properties going far beyond those announced in
\cite{AdS}, and we prove an additional locality property of such
theories on proper \AdS which was not observed in \cite{AdS}.
Moreover, we show that in two dimensions these suffice to construct a
nontrivial, local, covariant net indexed by bounded spacetime regions. 
We also explain how known examples of quantum fields on AdS fit into
our scheme.

     The primary lesson to be drawn from this paper is the observation that
covariance and passivity properties of states induce strong algebraic
relations between the observables, which may be interpreted as
manifestations of Einstein causality. In our research program, the 
theories on AdS treated here serve as a theoretical laboratory to 
test this striking feature. But the insight gained in this analysis 
goes beyond this class of field theoretical models to quantum fields
on other space--times. Further, it seems to be of relevance in the
discussion of causality problems appearing in nonlocal theories,
such as string theory and quantum field theory on noncommutative
space--times. 

\section{Unruh effect, PCT symmetry and weak locality} \label{main}
\setcounter{equation}{0} 

     We now enter into the analysis of the implications of our standing
assumptions (i)--(iv) by  appealing to a deep result of Pusz and 
Woronowicz for 
general quantum dynamical systems \cite{PuWo}. In the present context 
this result says that the vacuum vector $\Omega$ is, as a consequence of 
its passivity and mixing properties, invariant under the dynamics of any 
of the observers discussed above \cite[Theorem 1.1]{PuWo}.
In particular, this entails that $M_{01} \Omega = 0$ (and hence, by 
Lemma \ref{invariance}, $\Omega$ is invariant under the entire group 
$U(\AdSG)$), and $\omega$ is either \cite[Theorem 1.3]{PuWo} a ground 
state for $M_{01}$ (which is excluded by Lemma \ref{nogroundstate}), 
or satisfies, for some {\it a priori\/} unknown $\beta \geq 0$, 
the Kubo--Martin--Schwinger (KMS) condition. In fact, our assumptions 
exclude the possibility of $\beta = 0$. In the proof of Lemma 4.1 in 
\cite{PuWo} it is shown that if $\beta = 0$, then either $\omega$ is a 
trace state on $\As(W_R)$ or $M_{01} = 0$. In the second case, one
would have the triviality of the representation of the boost group 
and thus the triviality of $U(\AdSG)$, which is excluded by (i). 
The first case is excluded by assumption (iv). 
Therefore, for any pair of operators $A,B \in \As(W_R)$ 
there exists an analytic function $F$ in the strip 
$\{ z \in \CC \mid 0 < \Im(z) < \beta \}$ with continuous boundary values 
at $\Im(z) = 0$ and $\Im(z) = \beta$, which are given by
\begin{equation}  \label{kms}
F(t) = \omega( AB(t)) \, , \quad
F(t+i\beta) = \omega( B(t)A) \, ,
\end{equation}
respectively, for all $t \in \RR$ and with 
$B(t) \doteq e^{itM_{01}}Be^{-itM_{01}}$.
By the \AdSG--covariance the same assertions are valid for the action of the
groups $e^{itM_{0j}}$, $j = 2,\ldots,n-1$, on the suitable wedge algebras.
     
     In Appendix \ref{sectionReehSchlieder} it is proven that this
analyticity entails that the theories we are considering here 
satisfy the Reeh--Schlieder property (cf. Lemma \ref{ReehSchlieder}). 
So the vacuum vector $\Omega$ is cyclic for the algebra $\As(W)$, 
given any $W \in \Ws$, and, by the KMS-property, it is also separating 
for $\As(W)$ \cite[Corollary 5.3.9]{BratRob}. Hence, the 
Tomita--Takesaki modular theory is applicable to $(\As(W),\Omega)$, 
for every $W \in \Ws$ (cf. \cite{BR,KadRing}). Let $J_{W_R}$ denote the 
modular conjugation and $\Delta_{W_R}^{it}$ the modular unitaries 
associated to the pair $(\As(W_R),\Omega)$. 
Since the adjoint action of the strongly continuous unitary group
$e^{itM_{01}}$, $t \in \RR$, leaves the algebra $\As(W_R)$ invariant
and satisfies the KMS condition, we must have 
$\Delta_{W_R}^{it} = e^{-i\beta tM_{01}}$, for all $t \in \RR$  
\cite[Theorem 9.2.16]{KadRing}.  Hence, $J_{W_R}$ is determined by 
the equation
\begin{equation} \label{jwr}
J_{W_R} A \Omega = e^{\, -(\beta/2) M_{01}} A^* \Omega  
\, , \   A \in \CA(W_R) \, .
\end{equation}

\subsection{Unruh temperature} 

     The main task of this subsection is to determine the Unruh temperature
$\beta^{-1}$ and specific properties of the operator $J_{W_R}$. To this
end we shall adapt methods employed in \cite{BB}.

     We show in Lemma \ref{neighborhood} that there exists a wedge
$W_0 \in \Ws$ such that $\lambda W_0 \subset W_R$ for all $\lambda$ in
a neighborhood of the identity in $\AdSG$. Therefore, for any
$j = 2,\ldots,n-1$ one has
$\lambda_{0 j} (s) \, W_0 \subset W_R$ for the boosts $\lambda_{0 j} (s)$ 
in the $0$--$j$--plane for all sufficiently small parameters $s$. 
{}From equation (\ref{010j}) in Appendix \ref{grouptheory} we have   
\begin{equation} 
e^{\, itM_{0 1}} e^{\, isM_{0 j}} = 
e^{\, is(\cosh(t)M_{0 j} + \sinh(t) M_{1 j})}  e^{\, itM_{0 1}} \, .  
\end{equation}
Thus we get for any vector $\Phi \in \CH$ and operator $A^* \in \CA(W_0)$ 
\begin{equation} \label{equation11}
\langle \Phi , \, e^{\, itM_{0 1}} 
\, e^{\, isM_{0 j}} A^* e^{\, -isM_{0 j}}  \rva =  \langle \Phi , \, 
e^{\, is(\cosh(t)M_{0 j} + \sinh(t) M_{1 j})}\,  e^{\, itM_{0 1}} 
A^* \rva \, . 
\end{equation}
We are now in the position of employing the argument given in
\cite{AdS} to yield the equalities
\begin{equation} \label{equation13}
J_{W_R}  e^{\, isM_{0 j}}   =   
e^{\, is(\cos(\beta/2) M_{0 j} + i \sin(\beta/2) M_{1 j})} J_{W_R} \, .
\end{equation}
As pointed out in \cite{AdS}, the operator on the left--hand side 
of this equation is anti--unitary, which entails that $\beta$ is an 
{\it integer\/} multiple of $2 \pi$, for otherwise the operator 
appearing in the exponential function on the right--hand side would 
not be skew-adjoint. By using the proof of Theorem 6.2 in \cite{BB}
with $\As(\Os)$ replaced by $\As(W_0)$, one sees that its only 
possible value is $\beta = 2 \pi$. Proceeding to the proper time scale 
of the observer, we conclude that he is exposed to the Unruh 
temperature $(1/2\pi)((\dot{\lambda}_{0 1}(0)\, x_O)^2){}^{-1/2}$, 
in accordance with the value found in computations for 
some particular models \cite{DeLe,Jac} and also by more general 
considerations \cite{BEM}. 

    For geodesic observers, Lemma \ref{nogroundstate} is not applicable. 
In fact, $\omega$ cannot be a KMS-state for $e^{itM_{0n}}$ on 
$\As$, the laboratory observable algebra for geodesic observers. 
Indeed, since the covariance assumption (iii) implies 
$U(\lambda)\As U(\lambda)^{-1} = \As$,
for all $\lambda \in \AdSG$, if $e^{itM_{0n}}$ were the modular group
for $\Omega$ on $\As$, then modular theory would necessitate 
$U(\lambda)e^{itM_{0n}} = e^{itM_{0n}}U(\lambda)$, for all 
$\lambda \in \AdSG$ (cf. Thm. 3.2.18 in \cite{BR}). But this would 
only be possible if the representation $U(\AdSG)$ were trivial, which is 
excluded by assumption (i). So $\omega$ must be a ground state for
$e^{itM_{0n}}$.

     We have therefore established the following general facts:

\begin{theorem} \label{temperature}
Let Standing Assumptions (i)--(iv) hold. Then $\Omega$ is invariant under
the action of $U(\AdSG)$ and each uniformly accelerated observer testing 
$\Omega$  in \AdS finds a universal value 
$(1/2\pi)((\dot{\lambda}_{0 1}(0)\, x_O)^2){}^{-1/2}$ of the 
Unruh temperature which depends only on his particular orbit. 
For geodesic observers $\omega$ is a ground state; in particular, 
$M_{0n}$ is a positive operator.
\end{theorem}

\noindent This result is a consequence of the passivity of $\omega$,
and this vacuum state is the only normal state on $\As$ which 
is passive for all observers. In light of Theorem \ref{temperature}, 
it is physically justified to identify the operator $M_{0n}$ with the 
global energy operator. 

     The result $\beta = 2\pi$ and \cite[Theorem 9.2.16]{KadRing} 
permit us to completely determine the modular unitaries corresponding 
to the pair $(\As(W),\Omega)$, for all $W \in \Ws$.

\begin{cor} \label{modularunitary}
Given the Standing Assumptions (i)--(iv), the modular unitaries
for the pair $(\As(W_R),\Omega)$ are given by
\begin{equation}
\Delta_{W_R}^{it} = e^{-i2\pi tM_{01}} \, , \quad t \in \RR \, .
\end{equation}
\end{cor}

\noindent Covariance and the uniqueness of the modular objects yield
similar results for $\Delta_W^{it}$, for all $W \in \Ws$.

\subsection{PCT symmetry} 

     Having computed the modular unitaries and the value of the inverse 
temperature $\beta$ seen by the uniformly accelerated observers, let 
us return now to the analysis of $J_{W_R}$ and clarify its
relation to spacetime reflections. Plugging $\beta = 2 \pi$ into 
equation (\ref{equation13}), we see that for small $s$
\begin{equation}
J_{W_R}  e^{\, isM_{0 j}}   =   
e^{\, - is  M_{0 j}} J_{W_R} \, , \quad j = 2,\ldots,n-1 \, .
\end{equation}
This relation can be extended to arbitrary 
$s$ by iteration, if one decomposes $e^{\, isM_{0 j}}$ into an $m$--fold 
product $(e^{\, i(s/m)M_{0 j}})^m$ for sufficiently large $m$.
A similar argument with $\lambda_{0j}$ replaced by $\lambda_{0n}$
and equation (\ref{010j}) replaced by (\ref{010n}) yields
\begin{equation} \label{WR0n}
J_{W_R}  e^{\, isM_{0 n}}   =   
e^{\, - is  M_{0 n}} J_{W_R} \, .
\end{equation}
On the other hand, modular theory yields 
\begin{equation}
J_{W_R}  e^{\, isM_{0 1}}   =   
e^{\, is  M_{0 1}} J_{W_R} \, .
\end{equation}
Since these one-parameter subgroups generate $\AdSG$, the intertwining 
properties of $J_{W_R}$ with all unitaries $U(\lambda) \in U(\AdSG)$
are determined. 

\begin{lemma} \label{TCPgroup}
Given Standing Assumptions (i)--(iv), one has
\begin{equation}
J_{W_R} U(\lambda)  =  U(\theta_{01}\lambda\theta_{01}) J_{W_R} \, , 
\end{equation}
for all $\lambda \in \AdSG$, where $\theta_{01} = \diag(-1,-1,1,\ldots,1)$ 
is the reflection which changes the sign of the $0$--$1$--coordinates of 
the points in \AdS (the reflection about the edge of the wedge $W_R$). 
\end{lemma}

     Hence, if we define $U(\theta_{01}\lambda) \doteq J_{W_R}U(\lambda)$,
for any $\lambda \in \AdSG$, the following partial analogue of the 
PCT--theorem can be proven.

\begin{theorem} \label{proper}
If Standing Assumptions (i)--(iv) hold, then the unitary representation $U$ of 
\AdSG \, extends to a
representation of \AdSGG \, in which the reflection $\theta_{01}$
is implemented by the anti--unitary involution $J_{W_R}$.
\end{theorem}

\begin{proof} Note that $\AdSGG$ is the disjoint union of 
$\theta_{01}\AdSG$ and $\AdSG$. Since
\begin{equation}
U(\theta_{01} \lambda_1 \cdot \lambda_2) = J_{W_R} U(\lambda_1 \lambda_2)
= J_{W_R} U(\lambda_1)U(\lambda_2) = U(\theta_{01}\lambda_1)U(\lambda_2)
\end{equation}
and 
\begin{eqnarray}
U(\theta_{01} \lambda_1 \cdot \theta_{01}\lambda_2) & = &
U(\theta_{01} \lambda_1 \theta_{01})U(\lambda_2) \\ \nonumber
& = & J_{W_R} U(\lambda_1) J_{W_R} U(\lambda_2)
= U(\theta_{01}\lambda_1)U(\theta_{01}\lambda_2) \, ,
\end{eqnarray}
for all $\lambda_1,\lambda_2 \in \AdSG$, the assertion follows.
\end{proof}

\medskip

\noindent Theorem \ref{proper} is a purely group--theoretic statement which
does not yet say anything about the adjoint action of $J_{W_R}$ on the
observables. Results of that type require an additional assumption
and will appear in a later publication. 

\subsection{Weak locality} \label{weaklocalitysec}  

     In order to gain insight into the locality properties of the net, 
we consider the observables which are localized in the region
\begin{equation}
W_R{}^{\prime} \doteq \{x\in\AdS \mid - x_1 > | x_0 |, \, x_n > 0\} \, .
\end{equation}
Throughout this subsection we shall only consider 
\AdS of dimension $n \geq 3$.
Since the regions $W_R$ and $W_R{}^{\prime}$ are each one--half (on the
same ``side'' of $\AdS$ --- see Figure 1 below) of the regions 
obtained by intersecting opposite wedge--shaped regions in the ambient 
space $\RR^{n+1}$ with \AdS, we call them opposite wedges. In general, if
$W = \lambda W_R$, for some $\lambda \in \AdSG$, then $W' = \lambda W_R{}'$.

     In \cite{AdS} it was shown that for $n \geq 3$ the above results
entail that observables which are localized in complementary wedges
are weakly local in the vacuum state. So we can state:

\begin{theorem} \label{weaklocality1}
Under the assumptions (i)--(iv) and for $n \geq 3$, observables which are 
localized in opposite wedges of \AdS are weakly local with respect to each 
other. Explicitly, for any 
$\lambda \in \AdSG$ and any $A' \in \As(\lambda W_R{}')$, 
$B \in \As(\lambda W_R)$, one has 
$\langle\Omega,A'B\Omega\rangle = \langle\Omega,BA'\Omega\rangle$.
\end{theorem}

\noindent This result is also valid for theories in the covering space of
\AdS which satisfy (i)--(iv). A similar result was proven in \cite{BEM}
for two--point functions of quantum fields on the covering space of
\AdS satisfying a different set of assumptions in the Wightman framework.

     The uniqueness of the modular objects and the covariance
assumption (iii) give us a relation we shall use repeatedly in the
following:
\begin{equation} \label{unique}
U(\lambda) J_W U(\lambda)^{-1} = J_{\lambda W} \, ,
\end{equation}
for all $W \in \Ws$ and all $\lambda \in \AdSG$. Hence, since  
$\As(W_R{}') = e^{i\pi M_{12}}\As(W_R)e^{-i\pi M_{12}}$, 
equation (\ref{unique}) and the observations made above imply
\begin{equation}
J_{W_R{}'} = e^{i\pi M_{12}} J_{W_R} e^{-i\pi M_{12}} = 
J_{W_R} e^{-i2\pi M_{12}} = J_{W_R} \, .
\end{equation}
Thus, the $\AdSG$-covariance of the net entails
\begin{equation} \label{equation15}
J_{W'} = J_{W} \, , \quad \textnormal{for every $W \in \Ws$} \, .
\end{equation}
If the algebras $\As(W)$ and $\As(W')$ commuted strongly with each
other, it can be shown that (\ref{equation15}) follows from modular
theory. It is therefore of interest that (\ref{equation15}) obtains
when the algebras only weakly commute.

     We continue with some further locality results, which distinguish
theories on proper AdS from those on the covering space.  We
have chosen our wedge regions $W \in \Ws$ to be connected for
physical reasons, as previously explained. But the intersection of the
(connected) wedge {\bf W} in the ambient space
\begin{equation}
{\bf W} = \{ x \in \RR^{n+1} \mid x_1 > |x_0| \} 
\end{equation}
with \AdS has two connected components, one of which is $W_R$ and the
other is the conjugate wedge
\begin{equation}
\widetilde{W}_R{}' = \{ x \in \AdS \mid x_1 > |x_0|\, , -x_n > 0 \} \, .
\end{equation}
One has the geometric relations
\begin{figure}[b] \label{F1}
\epsfxsize98mm
\hspace*{40mm} \epsfbox{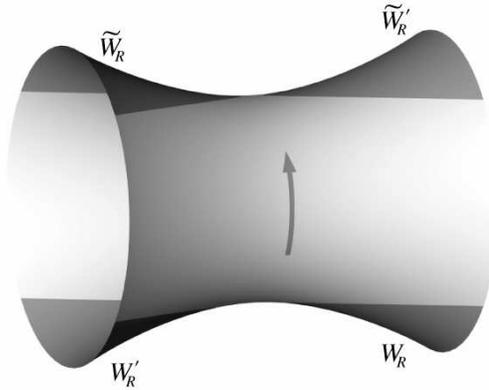}
\caption{Conjugate wedges in anti-de Sitter space and arrow of time.}
\end{figure}
\begin{equation}
\widetilde{W}_R = \theta_{01}\widetilde{W}_R{}^\prime = -W_R \quad 
\textrm{and} \quad W_R{}' = -\widetilde{W}_R{}' \, .
\end{equation}
Moreover, it is easy to see that
\begin{equation} \label{equation16}
e^{i\pi M_{0n}} \As(W_R) e^{-i\pi M_{0n}} = \As(\widetilde{W}_R{}') 
\quad \rm{and} \quad 
e^{i\pi M_{0n}} \As(W_R{}') e^{-i\pi M_{0n}} = \As(\widetilde{W}_R) \, .
\end{equation}
\noindent The rotation in the $0$--$n$--plane by $\pi$ also reverses 
the orientation
of the world lines $\lambda_{01}(t)x_O$ (cf. equation (\ref{0n01})), so that 
the world lines in $\widetilde{W}_R{}'$ run in the same direction as those 
of $W_R{}'$, while those in $\widetilde{W}_R$ have the same orientation 
as those in $W_R$. Explicitly, Corollary \ref{modularunitary} implies
that the modular unitaries for the pair $(\As(W_R{}'),\Omega)$ are given by

\begin{equation}
\Delta_{W_R{}'}^{it} = e^{i2\pi tM_{01}} \, , \quad t \in \RR \, .
\end{equation}
Hence, for all $t \in \RR$, one has
\begin{equation}
\Delta_{\widetilde{W}_R}^{it} = 
   e^{i\pi M_{0n}} \Delta_{W_R{}'}^{it} e^{-i\pi M_{0n}} = 
   e^{i\pi M_{0n}} e^{i2\pi t M_{01}} e^{-i\pi M_{0n}}
  = e^{-i2\pi t M_{01}} = \Delta_{W_R}^{it} \, ,
\end{equation}
by (\ref{0n01}).
So, $\{ e^{-i2\pi tM_{01}}\}_{t \in \RR}$ is the group of modular unitaries for
$(\As(\widetilde{W}_R),\Omega)$, and we have the relation
\begin{equation}
J_{\widetilde{W}_R} A \Omega = e^{\, -\pi M_{01}} A^* \Omega \ , \quad 
\mbox{for} \ \ A \in \CA(\widetilde{W}_R) \, .
\end{equation}
Moreover, we may appeal to (\ref{equation15}) and the modular theory to find
\begin{equation}
J_{\widetilde{W}_R} A' \Omega = e^{\, \pi M_{01}} A'{}^* \Omega \ , \quad 
\mbox{for} \ \ A' \in \CA(\widetilde{W}_R{}') \, .
\end{equation}
{}From equations (\ref{unique}) and (\ref{WR0n}) follow the equalities
\begin{equation} \label{0nR}
J_{\widetilde{W}_R{}'} = e^{i\pi M_{0n}} J_{W_R} e^{-i\pi M_{0n}} = 
e^{i2\pi M_{0n}}J_{W_R} = J_{W_R} 
\, ,
\end{equation}
where we have used $e^{i2\pi M_{0n}} = 1$, valid in proper AdS but not in
its covering space. Thus, by (\ref{equation15}) we have for $n \geq 3$:
\begin{equation} \label{Js}
J_{W_R{}'} = J_{W_R} = J_{\widetilde{W}_R} = J_{\widetilde{W}_R{}'} \, .
\end{equation}
We can now prove that $\As(\widetilde{W}_R{}')$ and $\As(W_R)$ are weakly 
local with respect to each other. 

\begin{theorem} \label{weaklocality2}
Under the assumptions (i)--(iv) and for $n \geq 3$, observables which are 
localized in $W_R$ are weakly local with respect to observables
localized in $\widetilde{W}_R{}'$. Explicitly, for any 
$\lambda \in \AdSG$ and any $A' \in \As(\lambda \widetilde{W}_R{}')$, 
$B \in \As(\lambda W_R )$, one has 
$\langle\Omega,A'B\Omega\rangle = \langle\Omega,BA'\Omega\rangle$.
\end{theorem}

\begin{proof} With the above preparations, one sees that for any
$A' \in \As(\widetilde{W}_R{}')$ and $B \in \As(W_R )$, 
one has
\begin{equation}
\begin{split}
\lva , A^{\prime *} B \rva & =  \overline{\lva , B^{*} A^{\prime}\rva}
= \overline{\lva , B^{*} J_{W_R} J_{W_R} A^{\prime}\rva}  \\ & = 
\lva , B e^{\, -\pi M_{0 1}} e^{\, \pi M_{0 1}} A^{\prime *} \rva 
=  \lva, B A^{\prime *} \rva \, .
\end{split}
\end{equation} 
\end{proof}

     Theorem \ref{weaklocality2} does not hold in the covering space
of \AdS, since, in general, $e^{i\pi M_{0n}}$ and $J_{W_R}$ will not
commute in such theories. Indeed, in \cite{F2,BEM} can be found examples
of a free field theory on the covering space of \AdS for which it can
be shown that assumptions (i)--(iv) are satisfied, but the elements of
$\As(W_R)$ and $\As(\widetilde{W}_R{}')$ do not commute in the vacuum
state --- see Appendix \ref{examples} for further discussion. However,
for theories on proper AdS, Theorem \ref{weaklocality2} is consistent
with a property of two-point functions observed in \cite{BEM} and
arrived at there by very different means and assumptions. 

     In Theorems \ref{weaklocality1} and \ref{weaklocality2} we see
that the passivity of the vacuum state and the group relations in
$\AdSG$ have determined which regions in \AdS are to have (weakly)
commensurable observables. In theories on AdS where the basic fields
satisfy standard c--number commutation relations, it follows from this result
that such observables actually commute in the usual (operator) sense 
\cite{StWi}. Indeed, it follows then that
\begin{equation} \label{locality}
\As(W_R ') \subset \As(W_R)' \quad \textnormal{and} \quad
\As(W_R) \subset \As(\widetilde{W}_R{}')' \, .  
\end{equation}

     But assumptions (i)--(iv) do not imply the strong 
locality relations (\ref{locality}) in general. Indeed, consider a 
tempered hermitian Fermi--type field $\phi$ with anticommutator
\begin{equation} \label{counterexample}
\phi(x)\phi(y) + \phi(y)\phi(x) = (W(x,y) + W(y,x)) \cdot 1 \, ,
\end{equation}
where $W(x,y)$ is taken to be any $\AdSG$--invariant two-point function
satisfying the spectrum condition. (Specific examples are provided
by two-point functions given in \cite[eq. (4.9)]{F2}). 
This anticommutator does not vanish when $x$ and $y$ are in 
complementary wedges. The field $\phi$ generates a CAR-algebra
with a quasifree state fixed by the two-point function
\begin{equation}
\omega( \phi(x) \phi(y)) = W(x,y) \, .
\end{equation}
Proceeding to the GNS representation with cyclic vector $\Omega$,
we conclude from \cite{F2,BEM} that this example satisfies all of 
our standing assumptions. Since $W(x,y)$ is symmetric when $x,y$ 
are in complementary wedges, the weak locality is explicit.

     As explained in \cite{AdS}, if (\ref{locality}) holds there are
grounds to expect that the theory generically cannot have
interaction. Theories which are weakly local but not strongly local
escape that argument.  In the following sections we shall
therefore continue to explore consequences of Standing Assumptions
(i)--(iv) when (\ref{locality}) does not hold, beginning with the
nature of the independence of the algebras $\As(W_1)$, $\As(W_2)$
associated with suitable spacelike separated wedges. But, first, we
make some further observations.
 
     Assumptions (i)--(iv) entail that each wedge algebra $\As(W)$ 
is a factor (cf. the proof of Theorem \ref{split}). If wedge
locality (\ref{locality}) held, then it would follow that 
$\As(W) \cap \As(W') = \CC 1$, {\it i.e.} 
no nontrivial observable can be localized in both $W$ and $W'$.  
It is therefore of physical interest that this fact also follows 
directly from (i)--(iv) in the absence of (\ref{locality}).

\begin{prop} \label{extendedlocality}
Let Standing Assumptions (i)--(iv) hold and let 
$n \geq 3$. Then for any
$W \in \Ws$, one has $\As(W) \cap \As(W') = \CC 1$ and 
$\As(W) \bigvee \As(W') = \Bs(\Hs)$.
\end{prop}

\begin{proof} By covariance, it suffices to prove the assertion for
$W = W_R$. Let $A \in \As(W_R) \cap \As(W_R{}')$ and $\Hs_0$
be the closure in $\Hs$ of $\big( \As(W_R) \cap \As(W_R{}') \big)\Omega$.
Since $e^{i2t\pi M_{01}}$, respectively $e^{-i2t\pi M_{01}}$, are the modular
unitaries corresponding to $(\As(W_R),\Omega)$, respectively 
$(\As(W_R{}'),\Omega)$,
then with $\Delta = e^{2\pi M_{01}}$ one has
\begin{equation}
J_{W_R} \Delta^{1/2} A \Omega = A^* \Omega \, , \quad  
J_{W_R{}'} \Delta^{-1/2} A \Omega = A^* \Omega \, .
\end{equation}
Hence, equation (\ref{Js}) entails that $\Delta^{1/2} = \Delta^{-1/2}$ on
$\Hs_0$. Therefore, $\Delta A \Omega = A \Omega$, which yields
$A \Omega = \Delta^{it} A \Omega = e^{i2t\pi M_{01}} A \Omega$, for
all $t \in \RR$. Hence, the Mean Ergodic Theorem (see, {\it e.g.}, 
\cite{Kato}) entails that $A \Omega = F_0 A \Omega$, where
$F_0$ is the projection onto the subspace of vectors in $\Hs$ each 
left invariant under $U(\AdSG)$, using Lemma \ref{invariance}.
Since the mixing property in condition (iii) entails that the rank of 
$F_0$ is 1 and since $\Omega$ is separating for $\As(W_R)$, it follows 
that $A$ is a multiple of the identity. But this entails
\begin{equation}
\CC 1 = J_W (\As(W) \cap \As(W')) J_W =
J_W \As(W) J_W \cap J_W \As(W') J_W =
\As(W)' \cap \As(W')' \, , 
\end{equation}
using (\ref{Js}), so that $\As(W) \bigvee \As(W') = \Bs(\Hs)$, for every 
$W \in \Ws$.
\end{proof} 

     Before we close this section, we have a final proposition to prove.

\begin{prop} \label{unequal}
Let Standing Assumptions (i)--(iv) hold, $n \geq 3$ and $W_1,W_2 \in \Ws$. If 
$W_2 \neq \pm W_1$, then $\As(W_1) \neq \As(W_2)$.
\end{prop}

\begin{proof}
Since two unequal wedges $W_1,W_2 \in \Ws$ have unequal reflections about 
their edges, unless $W_2$ coincides with $W_1{}', \widetilde{W}_1$ or 
$\widetilde{W}_1{}'$, Lemma \ref{TCPgroup} entails $J_{W_1} \neq J_{W_2}$
and, thus, $\As(W_1) \neq \As(W_2)$. 

     If $\As(W_1) \subset \As(W_1{}')$, then Theorem
\ref{weaklocality1} entails that the restriction of $\omega$ to
$\Rs(W_1)$ is a trace, which is excluded by assumptions (i) and
(iii). Similarly Theorem \ref{weaklocality2} yields $\As(W_1) \neq
\As(\widetilde{W}_1{}')$.
\end{proof}

\section{The Schlieder property} \label{Schliedersec}
\setcounter{equation}{0} 

     Many versions of the notion of independence of algebras of
observables in spacelike separated regions 
have emerged in algebraic quantum theory (see \cite{Sum}
for a review), and most are logically independent of the usual notion
of commensurability, which is that the algebras commute with each other
elementwise. In this section we shall prove that algebras associated with 
properly spacelike separated wedges $W_1,W_2$ satisfy an extended 
form of the algebraic
independence condition known as the Schlieder property, namely 
that $A_1 \in \As(W_1), A_2 \in \As(W_2)$ and $A_1 A_2 = 0$ imply 
either $A_1 = 0$ or $A_2 = 0$. We shall say that two wedges
$W_1,W_2$ are properly spacelike separated if $\lambda W_1 \subset W_2{}'$
for all $\lambda$ in some neighborhood of the identity in \AdSGG.
Note that $W$ and $W'$ are not properly spacelike separated. Although
in de Sitter and Minkowski spaces of dimension 
$n \geq 3$ such properly spacelike separated 
wedges do not exist, we show in Appendix \ref{sectioninclusion} that 
they are plentiful in AdS.

     The proof of an extended Schlieder property in \AdS, $n \geq 3$,
will be carried out in a series of steps.

\begin{lemma} \label{indprepare}
Let $W_1,W_2 \in \Ws$ be properly spacelike separated and let
$A_{1,k} \in \As(W_1)$, $A_{2,k} \in \As(W_2)$, $k = 1,\ldots,n$. 
If $B \in \Bs(\Hs)$ is such that
\begin{equation}
\sum_{k=1}^n A_{1,k} U(\lambda) B U(\lambda)^{-1} A_{2,k} = 0 \, , 
\end{equation}
for all $\lambda$ in some neighborhood $\Ns \subset \AdSG$ of the identity,
then this equality holds for all $\lambda \in \AdSG$.
\end{lemma}

\begin{proof} Choosing a smaller neighborhood $\Ns$, if necessary, it may
be assumed that there exists a $W \in \Ws$ such that
$W_1 \subset \lambda_0 \lambda_1 W$ and 
$W_2 \subset \lambda_0 \lambda_1 W^\prime$ for $\lambda_0, \lambda_1 \in \Ns$. 
Let $s \mapsto \lambda_W(s)$ be the group of 
boosts inducing a positive timelike flow on the wedge $W$ (and hence
a negative timelike flow on $W^\prime$). Setting
$\lambda_1(s) \doteq \lambda_1 \lambda_W(s) {\lambda_1}^{-1} $ 
for $\lambda_1 \in \Ns$, it will first be shown that
$B_s \doteq U(\lambda_1(s)) B U(\lambda_1(s))^{-1}$, $s \in \RR$, 
satisfies the hypothesis of the lemma as well. Putting
$\lambda(s) \doteq \lambda_0 \lambda_1 \lambda_W(s) 
{\lambda_1}^{-1} \lambda_0^{-1}$, $\lambda_0 \in N$,
and picking arbitrary elements $X_1$, $X_2$ in $\As(W_1)$, $\As(W_2)$, 
respectively, one has
\begin{equation}
\begin{split}
 \sum_{k=1}^n \lva, X_1 A_{1,k} \, U(\lambda_0) & B_s U(\lambda_0)^{-1} 
A_{2,k} X_2 \rva \\
  = \sum_{k=1}^n \lva, X_1 A_{1,k} \, 
U(\lambda(s)) U(\lambda_0) & B U(\lambda_0)^{-1} U(\lambda(s))^{-1} 
A_{2,k} X_2 \rva = 0 
\end{split}
\end{equation}
for sufficiently small $|s|$. 
Now $s \mapsto U(\lambda(s))$ is, after rescaling $s$, 
the modular group corresponding to 
$(\As(\lambda_0 \lambda_1 W),\Omega)$ and, similarly, 
$s \mapsto U(\lambda(s))^{-1}$ is the modular group corresponding
to  $(\As(\lambda_0 \lambda_1 W^\prime),\Omega)$. Since 
$\As(W_1) \subset \As(\lambda_0 \lambda_1 W)$ and
$\As(W_2) \subset \As(\lambda_0 \lambda_1 W^\prime)$, it follows that
\begin{equation}
s \mapsto \lva, X_1 A_{1,k} \, 
U(\lambda(s)) U(\lambda_0) B U(\lambda_0)^{-1} U(\lambda(s))^{-1} 
A_{2,k} X_2 \rva 
\end{equation}
extends to an analytic function on a strip of the upper complex half
plane for each $k = 1,\ldots,n$. By the preceding result, the 
corresponding sum of functions thus has to vanish for 
all $s \in \RR$. As $X_1$, $X_2$ were arbitrary within the above
limitations and $\Omega$ is cyclic for $ \As(W_1)$ and $ \As(W_2)$,
respectively, one concludes that 
$\sum_{k=1}^n A_{1,k} U(\lambda_0) B_s U(\lambda_0)^{-1} A_{2,k} = 0$, 
$s \in \RR$ and $\lambda_0 \in \Ns$.

     Next, let $\lambda_1,\ldots,\lambda_m \in \Ns$. Setting
$\lambda_i(s) \doteq \lambda_i \lambda_W(s) {\lambda_i}^{-1}$, 
$i = 1,\ldots,m$, one deduces by induction on $m$ that also
\begin{equation}
B_{s_1,\ldots,s_m} \doteq U(\lambda_m(s_m)) \cdots U(\lambda_1(s_1))
\, B \, U(\lambda_1(s_1))^{-1} \cdots U(\lambda_m(s_m))^{-1}
\end{equation}
satisfies the hypothesis of the lemma for $s_1,\ldots,s_m \in \RR$. 
Indeed, the case $m = 1$ has just been proven. By the induction
hypothesis and the group property of $U$, the assertion follows 
for $s_1,\ldots,s_{m-1} \in \RR$ and small $\vert s_m \vert$.
The argument presented in the preceding paragraph then entails
that the assertion holds for all $s_m \in \RR$.

But, according to Lemma \ref{generate}, the closure of 
the group generated by $\lambda_0 \lambda_W(s) {\lambda_0}^{-1}$,
$\lambda_0 \in \Ns$, $s \in \RR$, is \AdSG. Hence, taking into account
that $U$ is a continuous representation, it follows by (weak operator) 
continuity of $\lambda \mapsto U(\lambda) B U(\lambda)^{-1}$ that
these operators satisfy the hypothesis of the lemma as well, thereby 
completing its proof.
\end{proof} 

     In the following, we shall say that the wedge $W_1$ is properly
contained in the wedge $W_2$ and shall write $W_1 \Subset W_2$ if
there exists a neighborhood $\Ns$ of the origin in $\AdSG$ such that
$\lambda W_1 \subset W_2$, for all $\lambda \in \Ns$. Again, note that in
de Sitter space and Minkowski space of dimension $n \geq 3$
such pairs of wedges do not exist, but in
AdS they are abundant (see Appendix \ref{sectioninclusion}).

\begin{lemma} Let $W_1$, $W_2$ be properly spacelike separated, 
let $W$ be any wedge such that $W_1$, $W_2$ are properly contained in $W$
and $W^\prime$, respectively. If $A_{1,k}$, $A_{2,k}$, $k = 1,\ldots,n$,
are elements of $\As(W_1)$ and $\As(W_2)$, respectively, such that 
$\sum_{k=1}^n A_{1,k} A_{2,k} = 0$, one has 
\begin{equation}
\sum_{k=1}^n A_{1,k} U(\lambda_1) B_1 U(\lambda_1)^{-1} \cdots 
U(\lambda_m) B_m U(\lambda_m)^{-1} A_{2,k} = 0 \, ,
\end{equation}
for $m \in \IN$ and $B_i \in \As(W)'$, 
$\lambda_i \in \AdSG$, $i=1,\ldots,m$.
\end{lemma}

\begin{proof} The proof of the lemma proceeds by induction on $m$. 
Let $\Ns$ be a neighbourhood of the identity in \AdSG \, such that 
${\lambda_0}^{-1} W_1 \subset W$ for $\lambda_0 \in \Ns$. Then
\begin{equation}
U(\lambda_0)^{-1} A_{1,k} U(\lambda_0) \in \As({\lambda_0}^{-1} W_1)
\subset \As(W) = (J_W \As(W) J_W)^\prime  \, .
\end{equation}
So $A_{1,k}$ and $U(\lambda_0) B_0 U(\lambda_0)^{-1}$ commute for
any $B_0 \in J_W \As(W) J_W$ and consequently 
\begin{equation}
0 = U(\lambda_0) B_0 U(\lambda_0)^{-1} \sum_{k=1}^n A_{1,k} A_{2,k} =
\sum_{k=1}^n A_{1,k} U(\lambda_0) B_0 U(\lambda_0)^{-1} A_{2,k} \, . 
\end{equation}
By the preceding lemma, this equality
extends to all $\lambda_0 \in \AdSG$. Assuming now that the
statement holds for $m$, one has with the same choices of 
$B_0$ and $\lambda_0$ as in the preceding step
\begin{equation}
\begin{split}
0 & = \sum_{k=1}^n U(\lambda_0) B_0 U(\lambda_0)^{-1} 
A_{1,k} U(\lambda_1) B_1 U(\lambda_1)^{-1} \cdots 
U(\lambda_m) B_m U(\lambda_m)^{-1} A_{2,k}  \\
& = \sum_{k=1}^n A_{1,k} U(\lambda_0) B_0 U(\lambda_0)^{-1}  
U(\lambda_1) B_1 U(\lambda_1)^{-1} \cdots 
U(\lambda_m) B_m U(\lambda_m)^{-1} A_{2,k} \, , 
\end{split}
\end{equation}
for  $B_i \in J_W \As(W) J_W$, $\lambda_i \in \AdSG$, $i=1,\ldots,m$.
Taking into account that $U$ is a representation of \AdSG, this
implies (after an obvious redefinition of $\lambda_1,\ldots,\lambda_m$)
\begin{equation} 
\sum_{k=1}^n A_{1,k} U(\lambda_0) \big( B_0 U(\lambda_1) B_1 U(\lambda_1)^{-1} 
\cdots U(\lambda_m) B_m U(\lambda_m)^{-1} \big) U(\lambda_0)^{-1} A_{2,k} 
= 0 \, .
\end{equation}
Applying once more the preceding lemma, one concludes that
\begin{equation}
\sum_{k=1}^n A_{1,k} U(\lambda_0) B_0 U(\lambda_0)^{-1}  
U(\lambda_1) B_1 U(\lambda_1)^{-1} \cdots 
U(\lambda_m) B_m U(\lambda_m)^{-1} A_{2,k} = 0 \, ,
\end{equation}
for $B_i \in J \As(W) J$, $\lambda_i \in \AdSG$, $i=0,1,\ldots,m$,
completing the proof.
\end{proof}

     In the next step of our argument we make use of the relation
$J_W U(\lambda) J_W = U(\theta \lambda \theta)$, where $\theta$ is the
reflection about the edge of $W$ (cf. Lemma \ref{TCPgroup}). Because of 
weak additivity, Proposition \ref{extendedlocality} and the preceding
relation we have
\begin{equation}
\begin{split}
\bigvee_{\lambda} U(\lambda) J_W \As(W) J_W U(\lambda)^{-1}
& = J_W \Big( \bigvee_{\lambda} U(\lambda) \As(W)
U(\lambda)^{-1} \Big) J_W \\ & = J_W \Bs(\Hs) J_W  = \Bs(\Hs) \, ,
\end{split}
\end{equation}
showing that $\Bs(\Hs)$ is the weak operator closure of the algebra 
generated by the operators $U(\lambda) J_W B J_W U(\lambda)^{-1}$, 
$B \in \As(W)$, $\lambda \in \AdSG$. It therefore follows 
that for any collection of operators $A_{1,k}$, $A_{2,k}$ as in the preceding 
lemma one has 
\begin{equation}
\sum_{k=1}^n A_{1,k} B A_{2,k} = 0 \, , \quad B \in \Bs(\Hs) \, .
\end{equation}
Taking into account that $\Bs(\Hs)$ contains in particular all operators
of rank $1$, we conclude that for any normal state $\tilde{\omega}$
on $\Bs(\Hs)$ we have
\begin{equation}  \label{Schlieder}
\sum_{k=1}^n \tilde{\omega}(A_{1,k})A_{2,k} = 0 = 
\sum_{k=1}^n A_{1,k} \tilde{\omega}(A_{2,k}) \, .   
\end{equation}
So we have established that properly spacelike separated wedge
algebras manifest a strong form of algebraic independence which implies the 
Schlieder property.

\begin{prop} Let Standing Assumptions (i)--(iv) hold and let
$W_1$ and $W_2$ be properly spacelike separated wedges. For any 
$A_{1,k} \in \As(W_1)$ and $A_{2,k} \in \As(W_2)$,
$k = 1,\ldots,n$, such that $\sum_{k=1}^n A_{1,k} A_{2,k} = 0$, 
relation (\ref{Schlieder}) must hold for all normal states 
$\tilde{\omega}$ on $\Bs(\Hs)$. In particular, if $A_1 A_2 = 0$, 
then either $A_{1} = 0$ or $A_{2} = 0$.
\end{prop} 

     If the algebras $\As(W_1)$ and $\As(W_2)$ were mutually
commuting, then the Schlieder property is equivalent to
$C^*$-independence \cite{Roos}. However, in the noncommuting case, the
Schlieder condition is strictly weaker than $C^*$-independence
\cite{Ham0}, and it is an open question in our setting whether
$C^*$-independence holds if $\As(W_1)$ and $\As(W_2)$ do not commute.

\section{The split property} \label{splitsec}
\setcounter{equation}{0} 

     We shall next show that if $W_1$ is properly contained in $W_2$,
then there exists a type I factor $\Ms$ such that $\As(W_1) \subset
\Ms \subset \As(W_2)$, as long as the multiplicity of the eigenvalues
of $M_{0n}$ does not grow too fast. Hence, with this additional
assumption, the algebras $\As(W_1)$ and $\As(W_2)'$ manifest a
particularly strong form of statistical independence. 

     Since $\AdS$ is periodic in the time variable and $M_{0n}$ is a 
positive operator, the spectrum of $M_{0n}$ is a subset of 
$\IN_0 = \{ 0 \} \cup \IN$. 
If the multiplicities of the eigenvalues of $M_{0n}$
do not increase too rapidly, then $e^{-\gamma M_{0n}}$ is a trace 
class operator for any $\gamma > 0$. In Appendix \ref{multiplicity} 
we exhibit simple examples, constructed from
irreducible unitary positive energy representations of $\AdSG$,
where this situation obtains.
We formulate this assumption explicitly as condition (NC).

\medskip

  (NC) \, There exist constants $c_0 > 0$ and $0 < k_0 < 1$
such that the spectral multiplicities $\boldsymbol{\mu}_m$ 
of the eigenvalues $m$ of $M_{0n}$ are bounded by 
$\boldsymbol{\mu}_m \leq e^{c_0 \, m^{k_0}}$, $m \in \IN_0$.

\medskip
     It is particularly straightforward to establish the 
``split property'' in the presence of condition (NC). 
In fact, in \cite[Thm. 3.2]{DLR} it was shown that in conformally 
invariant theories a trace-class condition on the exponentiated 
conformal Hamiltonian entails that suitable inclusions are split.
We indicate here a somewhat
different and more explicit argument. To this end, we recall the 
following lemma \cite[Lemma 2.3]{BDF}. 
\begin{lemma} Let $U(t) = e^{itH}$, $t \in \RR$,  determine
a strongly continuous one-parameter group of unitary operators 
with positive generator $H$ 
and invariant unit vector
$\Omega \in \Hs$. Moreover, let 
$\As$ and $\Bs$ be von Neumann algebras satisfying
\begin{equation} \label{comm}
U(t) \As U(t)^{-1} \subset \Bs \, , 
\end{equation}
for all $\vert t \vert < \delta$ and some $\delta > 0$. Then there  
exists a continuous function
$f : \RR \rightarrow \RR$ which decreases almost exponentially, {\it i.e.}
$\sup_\omega \vert f(\omega) \vert e^{\vert \omega \vert^k} < \infty$, for
any $0 < k < 1$, such that
\begin{equation}
\lva,AB'\rva = \lva,Af(H)B'\rva + \lva,B'f(H)\rva \, ,  
\end{equation}
for all $A \in \As$ and $B' \in \Bs '$. \label{bdf}
\end{lemma}

     We use this to establish the following general result.
     
\begin{prop} \label{splitabstract}
Let $U(t) = e^{itH}$, $t \in \RR$,  be 
a strongly continuous one-parameter group of unitary operators 
with an (up to a phase) unique invariant unit vector $\Omega \in \Hs$ and 
with a generator $H$ having spectrum in $\IN_0$ 
and spectral multiplicities 
bounded by $\boldsymbol{\mu}_m \leq e^{c_0 \, m^{k_0}}$, $m \in \IN_0$,
for fixed $c_0 > 0$ and $0 < k_0 < 1$. 
Moreover, let $\As$ and $\Bs$ be type III factors
with $\Omega$ cyclic and separating for both and satisfying
\begin{equation} \label{comm1}
U(t) \As U(t)^{-1} \subset \Bs \, , 
\end{equation}
for all $\vert t \vert < \delta$ and some $\delta > 0$. Then there 
exists a type $I$ factor $\Ms$ such that 
$\As \subset \Ms \subset \Bs$. 
\end{prop}
\begin{proof} Consider the algebraic
tensor product $\As \odot \Bs'$ and two of its representations

   (a) $(\pi,\Hs,\Omega)$ with $\pi(A \odot B') \doteq A \, B'$

   (b) $(\pi_p,\Hs \otimes \Hs,\Omega\otimes\Omega)$ with 
$\pi_p(A \odot B') \doteq A \otimes B'$ \,

\noindent in an obvious notation. By the preceding lemma, one has 
\begin{equation}
\lva,\pi(A \odot B')\rva = \lva,AB' \rva =
\lva,Af(H)B'\rva + \lva,B'f(H)\rva \, . \label{zip}
\end{equation}
Let $P_\Omega$ be the projection onto $\CC\Omega \subset \Hs$ and let
$V$ be the unitary flip on $\Hs\otimes\Hs$:
\begin{equation}
V \Phi \otimes \Psi \doteq \Psi \otimes \Phi \, , \quad 
\Phi,\Psi \in \Hs \, .
\end{equation}
Taking into account that $f(H)$ is a trace class operator on $\Hs$,
because of the assumptions on the spectrum of $H$ and the rapid decay 
of $f$, one may conclude from (\ref{zip}) that
\begin{equation}
\lva,\pi(A \odot B')\rva =  tr_{\Hs\otimes\Hs}  \left(
\left[ V(f(H) \otimes P_\Omega) + (P_\Omega \otimes f(H))V \right]
A \otimes B' \right) \, ,
\end{equation}
where the operator in square brackets is of trace class on 
$\Hs\otimes\Hs$. Thus,
\begin{equation}
\lva,\pi(A \odot B')\rva = \omega_p(\pi_p (A \odot B')) \, , 
\end{equation}
where $\omega_p$ is some normal functional with respect to the representation
$\pi_p$. Since the left hand side defines a state on $\As\odot\Bs'$,
so does the right hand side; hence, $\omega_p$ is, in fact, a state
on $\pi_p(\As\odot\Bs')^- = \As \overline{\otimes} \Bs'$. Since 
$\As,\Bs'$ are type III factors, so is their tensor product. 
Moreover, $\Omega \otimes \Omega$ is a cyclic and separating vector for
$\As \overline{\otimes} \Bs'$, since $\Omega$ is cyclic and 
separating for $\As$ and $\Bs'$. Hence $\omega_p$ is represented by a
vector $\Omega_p \in \Hs \otimes \Hs$, and one has
\begin{equation}
\lva,\pi(A \odot B')\rva = \lvap,\pi_p (A \odot B')\rvap \, ,
\quad A \in \As, B' \in \Bs' \, . 
\end{equation}
Since $\Omega$ is cyclic for $\pi(\As\otimes\Bs')$, one concludes that
$\pi$ is unitarily equivalent to some subrepresentation of $\pi_p$.
But, due to the fact that $\pi_p(\As\otimes\Bs')$ is a factor of
type III, any subrepresentation of $\pi_p$ is equivalent to 
$\pi_p$; hence, $\pi$ and $\pi_p$ are unitarily equivalent. 

     So let $W : \Hs\otimes\Hs \rightarrow \Hs$ be a unitary such that
\begin{equation}
AB' = \pi(A \odot B') = W\pi_p(A \odot B')W^{-1} = W A \otimes B'\,  W^{-1} \,
. 
\end{equation}
Since $\As \otimes 1 \subset \Bs(\Hs) \otimes 1 \subset (1 \otimes \Bs')'$,
one concludes, after applying the adjoint action of $W$ to this chain
of inclusions,
\begin{equation}
\As \subset W (\Bs(\Hs) \otimes 1) W^{-1} \subset \Bs'' = \Bs \, ,
\end{equation}
where $\Ms \doteq W (\Bs(\Hs) \otimes 1) W^{-1}$ is a type I factor.
\end{proof}

     We can now prove the following theorem for our immediate purposes.

\begin{theorem} \label{split}
Let the assumptions (i)--(iv) and (NC) hold. Then for any wedges 
$W_1$ and $W_2$ such that
\begin{equation}  \label{propercontain}
e^{itM_{0n}} \As(W_1) e^{-itM_{0n}} \subset \As(W_2) \, ,
\end{equation}
for all sufficiently small $t \in \RR$, there exists a type $\rm{I}_{\infty}$
factor $\Ms$ such that $\As(W_1) \subset \Ms \subset \As(W_2)$.
\end{theorem}

\noindent {\bf Remark}: Of course, if $W_1 \Subset W_2$, then 
(\ref{propercontain}) holds.

\begin{proof} Our passivity and mixing assumptions entail that $\As(W)$
is a type $\rm{III}_1$ factor, for all $W \in \Ws$ \cite[Theorem 4.3]{PuWo}.
The theorem then follows at once from Proposition \ref{splitabstract}.
\end{proof}

     Note that in de Sitter space and Minkowski space of 
dimension $n \geq 3$ no inclusions 
of wedge algebras can be split. For Minkowski space
this was observed in \cite{Bu}; in de Sitter space $W_1 \subset W_2$
entails $W_1 = W_2$.
 
     The above results hold only for theories on proper AdS. If the
covering space of AdS is considered, then condition (NC) must be 
replaced by the condition that the map
\begin{equation}
\As(W) \ni A \mapsto e^{-\gamma M_{0n}}A\Omega \, , 
\end{equation}
is nuclear, for all $\gamma > 0$ \cite{BuWi}. The proof presented 
in \cite{BuWi} is formulated in terms of double cone algebras in 
Minkowski space but carries over to the present situation without
difficulty.

\section{Local nets on two-dimensional AdS} \label{2d}
\setcounter{equation}{0} 

     In this section we prove that Standing Assumptions (i)--(iv)
and the spectral condition (NC) entail the existence of a nontrivial, 
covariant and local subnet in two-dimensional AdS. It is noteworthy that 
locality properties can be derived in these circumstances, and
it would be of interest to see whether the  
same is also true in higher-dimensional AdS.

     In two-dimensional AdS, the set of all wedge-shaped regions 
consists of two disconnected pieces 
$\{ \lambda W_R \mid \lambda \in \AdSG \}$ and
$\{ \lambda W_L \mid \lambda \in \AdSG \}$, where $W_L = W_R{}'$. 
The former (respectively, latter) will be denoted by $\Ws_R$ ($\Ws_L$) and 
called the class of right wedges (left wedges). Note that 
$W \in \Ws_R$ if and only if $W' \in \Ws_L$. We assume as before that
\begin{equation}  \label{leftweakadditivity}
\As = \bigvee_{W \in \Ws_R} \As(W) \, ,   
\end{equation}
and that $\Omega$ is cyclic for $\As$. Hence, once again, 
Proposition \ref{ReehSchlieder} yields the cyclicity of $\Omega$ for 
every $\As(W)$, $W \in \Ws_R$.

     In $\AdSt$ the edge of a wedge is a single point. For each 
point $a \in \AdSt$ we shall denote by $W_a$ the unique element 
of $\Ws_R$ whose edge is $a$. Then $W_a{}'$ is the unique element 
of $\Ws_L$ whose edge is $a$. For $a,b \in \AdSt$ such that
$W_b \Subset W_a$ 
we define the open sets (double cones) $\Os_{a,b} = W_a \cap W_b{}'$. 
Corresponding to $\Os_{a,b}$ we define the von Neumann algebra
\begin{equation}
\Bs(\Os_{a,b}) = \As(W_a) \cap \As(W_b)' \, . 
\end{equation}
Note that $\Os_{a,b} \subset \Os_{c,d}$ if and only if $W_a \subset W_c$
and $W_d \subset W_b$. The isotony of the original net of wedge algebras
then implies
\begin{equation}
\Bs(\Os_{a,b}) \subset \Bs(\Os_{c,d}) \, , 
\end{equation}
\ie the isotony of the net $\{ \Bs(\Os_{a,b}) \mid a,b \in \AdSt \}$.
The covariance of the original net of wedge algebras entails
\begin{equation}
\begin{split}
U(\lambda)\Bs(\Os_{a,b})U(\lambda)^{-1}  & =  
  U(\lambda)(\As(W_a) \cap \As(W_b)')U(\lambda)^{-1} \\ & = 
  \As(\lambda W_a) \cap \As(\lambda W_b)' 
  =  \Bs(\lambda \Os_{a,b}) \, 
\end{split}
\end{equation}
for all $\lambda \in \AdSGG$. 

     In general, there is no reason for such relative commutants 
of wedge algebras to be nontrivial. But in our setting they turn
out to be very large algebras, having a dense $G_\delta$ set of cyclic vectors.

\begin{prop} If conditions (i)--(iv) and (NC) 
hold in a theory on two-dimensional AdS and $W_b \Subset W_a$, 
then $\Bs(\Os_{a,b})$ is a type $\rm{III}_1$ factor.
\end{prop}

\begin{proof} If $W_b$ is properly contained in $W_a$, then from 
Theorem \ref{split} one has the existence of a type I factor $\Ms$ such that 
$\As(W_b) \subset \Ms \subset \As(W_a)$. {}From the proof of Theorem 2.1
in \cite{DopLon} there exists a unitary mapping 
$\Hs \rightarrow \Hs \otimes \Hs$ such that 
$U \As(W_b) U^{-1} = \As(W_b) \otimes 1$ and 
$U \As(W_a) U^{-1} = \Bs(\Hs) \otimes \As(W_a)$. Using Takesaki's commutant
theorem for tensor products (cf. \cite[Theorem 11.2.16]{KadRing}), one 
therefore sees that
\begin{equation}
\begin{split}
\Bs(\Os_{a,b}) & =  \As(W_a) \cap \As(W_b)' =
U^{-1} \big( (\Bs(\Hs) \otimes \As(W_a)) \cap (\As(W_b)' 
\otimes \Bs(\Hs))\big)U
\\
 & =  U^{-1} \As(W_b)' \otimes \As(W_a) U \, .
\end{split}
\end{equation}
Since $\As(W_a)$ and $\As(W_b)$ are type $\rm{III}_1$ factors (cf. proof of 
Theorem \ref{split}), the algebra $\As(W_b)' \otimes \As(W_a)$ and, thus, 
also $\Bs(\Os_{a,b})$ is a type $\rm{III}_1$ factor (cf. 
\cite[Corollary 11.2.17]{KadRing} and 
\cite[Th\'eor\`emes 1.3.4 and 3.4.1]{Con}).
\end{proof}

     We have seen previously that assumptions (i)--(iv) prescribe the
sense in which ``spacelike separated'' is to be understood in \AdS.
Hence, the region $\Os_{a,b}$ is spacelike separated from $\Os_{c,d}$
if there exists a wedge $W$ such that $\Os_{a,b} \subset W$
and $\Os_{c,d} \subset W'$. Without loss of generality, we may assume
for concreteness that $W \in \Ws_R$, so that $W_a \subset W \subset W_d$. 
Then one finds that the local algebras satisfy locality:
\begin{equation}
\begin{split}
\Bs(\Os_{a,b}) & =  \As(W_a) \cap \As(W_b)' \subset \As(W)
\subset \As(W_c)' \vee \As(W_d) =  \Bs(\Os_{c,d})' \, .
\end{split}
\end{equation}

     We summarize these findings in the following theorem.

\begin{theorem} \label{localnet} 
If conditions (i)--(iv), (NC) and (\ref{leftweakadditivity}) hold in a
theory on two-dimensional AdS, then the above construction yields
a nontrivial, covariant and local net $\Os \mapsto \Bs(\Os )$ based on
double cones  
$\Os \subset \AdSt$, in which each algebra $\Bs(\Os)$ is a factor of
type $\rm{III}_1$. 
\end{theorem}

\noindent One can similarly define a second nontrivial, covariant and
local net from a given net of wedge algebras  
based upon the wedges in $\Ws_L$ . These two local nets 
coincide if the initial nets of wedge algebras 
are local with respect to each
other, $\As(W^\prime) \subset \As(W)^\prime$.

     We wish to sketch some consequences of this construction. 
The weakly local, but quite nonlocal net fixed by the field $\phi$ given in 
equation (\ref{counterexample}) is also well-defined in
two-dimensional AdS, as long as the two-point function $W(x,y)$ 
is suitably chosen. The above construction yields many nontrivial
observables localized in precompact subsets of $\AdSt$ and 
associated with this field. In terms of the original, simply 
expressed field, these local observables are quite complicated
objects. This suggests the possibility of constructing complex
local objects from relatively simple nonlocal fields.

     In \cite{BL} the existence of split inclusions of wedge algebras
was replaced by a modular nuclearity condition to employ the above
construction in theories on two-dimensional Minkowski space.
This modular nuclearity condition has been verified \cite{L} in a 
factorizing S-matrix model with S-matrix not equal to the identity.
The basic field in that model has algebraic relations similar to
those of our field $\phi$. Since in some suitable sense quantum
theories on AdS become close to theories on Minkowski space
as the AdS radius becomes sufficiently large, one may expect that
the model on $\AdSt$ determined by $\phi$ also describes physics
which goes beyond that of generalized free fields. These matters shall be
further investigated elsewhere.

\section{Conclusions and further remarks}
\setcounter{equation}{0} 

     We have thus shown that stability properties of a state carrying
the interpretation of a ``vacuum'' imply a PCT theorem, the uniqueness
of the Unruh temperature, as well as commensurability and
independence properties of the observables in any quantum field theory
on AdS. Such implications exist also in other space--times, but they
are of particular interest in the case of AdS, where the causal
structure is such that it is not clear {\it a priori} how to define
``mutually spacelike regions,'' and therefore it is not clear which
locality relations are physically meaningful. Nor is it sufficient to
sidestep the issue by appealing to theories on the covering space of
\AdS. In fact, we have seen that observables in opposite wedges $W$
and $W^{\prime}$ or in conjugate wedges $W$ and $\widetilde{W}'$
necessarily commute with each other, either weakly or strongly. The
former fact can seem natural from the point of view of the covering
space of \AdS, but the latter can certainly not be observed in
theories on the covering space. It is remarkable that locality
properties can be determined by stability assumptions. Indeed, we showed
that in two-dimensional AdS these stability assumptions and a weak
growth condition on the eigenvalues of $M_{02}$ entail the existence
of a nontrivial, covariant, local net indexed by bounded regions in
AdS.  It would be of interest to determine if the same is true of
theories on higher dimensional AdS.

     It was conjectured by Bros, Epstein and Moschella in \cite{BEM} 
that the assumptions made in that paper should follow from our standing 
assumptions. This would be an interesting matter to settle, since
it would allow one to derive more detailed
information about the analyticity properties of the two-point function
of quantum fields on AdS.

\bigskip

\noindent {\bf Acknowledgements}: DB and SJS wish to thank the
Department of Mathematics and the Institute for Fundamental Theory of
the University of Florida and the Institute for Theoretical Physics at
the University of G\"ottingen, respectively, for hospitality and
financial support. DB is also grateful for financial support by the
Deutsche Forschungsgemeinschaft (DFG). We have benefitted from
discussions with J. Bros. We also acknowledge
conversations with M. Florig at an early stage of these
investigations.  \\

\appendix

\section{Unitary representations of the AdS group}  \label{grouptheory}
\setcounter{equation}{0} 

     The algebraic properties of the strongly continuous unitary 
representation of \AdSG \, assumed to exist in (i) are of crucial
importance to us. As a convenience to the reader, we collect the relevant
properties here. Many of the arguments in this section are adapted from
\cite{BB}. Given any such representation $U$ and a coordinate system
on \AdS, we denote by $M_{\mu\nu}$, $\mu,\nu = 0,1,\ldots,n$, the 
corresponding self--adjoint generators. On a dense, invariant domain of 
analytic vectors in $\Hs$, they fulfill the following 
Lie-algebraic relations.
\begin{equation} \label{liealgebra}
[M_{\mu\nu},M_{\rho\sigma}] = -ig_{\mu\rho}M_{\nu\sigma} +
ig_{\mu\sigma}M_{\nu\rho} - ig_{\nu\sigma}M_{\mu\rho} + ig_{\nu\rho}M_{\mu\sigma} \, ,
\end{equation}
where $g = \diag(1,-1,\cdots,-1,1)$, $\mu,\nu = 0,1,\ldots,n$.
In particular, the operator $M_{01}$ generates the action of the boosts
$\lambda_{01}$,
\begin{equation}
U(\lambda_{01}(t)) = e^{itM_{01}} \, , \quad t \in \RR \, ,
\end{equation}
and similarly for the operators $M_{0j}$, $j = 2,\ldots,n-1$. $M_{0n}$ is 
the generator of the time translations. Any of the operators 
$M_{0j}$, $j = 1,\ldots,n$, may be taken to be the operator $M$ discussed 
in the Introduction (see Section \ref{intro}). If $n \geq 3$, the operators 
$M_{jk}$, $j,k = 1,\ldots,n-1$, are the generators of spatial rotations, 
whereas $M_{jn}$, $j = 2,\ldots,n-1$ generate other subgroups of boosts
obtained from the first-mentioned by temporal rotation. If $n = 2$, 
then $M_{01}$ and $M_{12}$ are generators of boosts and there are no 
spatial rotations.

     The Lie-algebraic relations (\ref{liealgebra}) yield the equality
\begin{equation}
e^{isM_{0n}}M_{01}e^{-isM_{0n}} = \cos(s) M_{01} - \sin(s) M_{1n}
\, ,
\end{equation}
which implies
\begin{equation}
e^{i\pi M_{0n}}M_{01}e^{-i\pi M_{0n}} = - M_{01} \, .
\end{equation}
Hence, $M_{01}$ cannot be a positive operator on $\Hs$. Since the 
representation $U(\AdSG)$ is nontrivial, we may conclude the following 
lemma from the assumed $\AdSG$--covariance.

\begin{lemma} \label{nogroundstate}
The operators $M_{0j}$, $j = 1,\ldots,n-1$, cannot be positive 
on $\Hs$ and thus $\Omega$ cannot be a ground state for the group
$e^{itM_{0j}}$.
\end{lemma}

\noindent  The Lie-algebraic relations (\ref{liealgebra}) also imply 
the following group relations:
\begin{eqnarray} \label{0n01}
& e^{isM_{0n}}e^{itM_{01}}e^{-isM_{0n}} = 
e^{it(\cos(s) M_{01} - \sin(s) M_{1n})} \, ; & \\
\label{equation6}
& e^{isM_{0j}}e^{itM_{01}}e^{-isM_{0j}} = 
e^{it(\cosh(s) M_{01} - \sinh(s) M_{1j})} \, , 
\quad j = 2,\ldots,n-1 \, ; & \\
\label{equation7}
& e^{isM_{j1}}e^{itM_{01}}e^{-isM_{j1}} = 
e^{it(\cos(s) M_{01} - \sin(s) M_{j0})}
\, , \quad j = 2,\ldots,n-1 \, ; & \\
\label{010n}
& e^{isM_{01}}e^{itM_{0n}}e^{-isM_{01}} = 
e^{it(\cosh(s) M_{0n} - \sinh(s) M_{1n})} \, . &
\end{eqnarray}

\noindent Of course, equations (\ref{equation6}) and (\ref{equation7})
are vacuous if $n = 2$. We shall establish a few necessary tools.

\begin{lemma} \label{generate}
Let $\Ns \subset \AdSG$ be any neighborhood of the identity
in \AdSG. Then the closure $\Us_{\Ns}$ in the strong operator topology
of the group generated by the 
unitaries $U(\lambda \lambda_{01}(t)\lambda^{-1})$, $t \in \RR$, 
$\lambda \in \Ns$, contains $U(\AdSG)$.
\end{lemma}

\begin{proof} Consider first $n \geq 3$. {}From (\ref{equation6}) it follows 
that for all sufficiently small 
$\vert s \vert$ and for all $t \in \RR$, $\Us_{\Ns}$ contains the operator
$e^{it(\cosh(s) M_{01} - \sinh(s) M_{1j})}$ for $j = 2,\ldots,n-1$. With
fixed $s$, the Trotter product formula \cite{RS} applied to the product of
the one--parameter groups $e^{it(\cosh(s) M_{01} - \sinh(s) M_{1j})}$ and
$e^{it\cosh(s) M_{01}} \in \Us_{\Ns}$ implies that the rotations
$e^{-it\sinh(s) M_{1j}}$, $j = 2,\ldots,n-1$, are also contained in
$\Us_{\Ns}$. Similarly, equations (\ref{0n01}) and (\ref{equation7}) entail 
that the operators $e^{itM_{1n}}$ and $e^{itM_{j0}}$, $j = 2,\ldots,n-1$, 
$t \in \RR$, are contained in $\Us_{\Ns}$. Since the group $U(\AdSG)$ is 
generated by these subgroups, the proof is complete for $n \geq 3$.

     If $n = 2$, then the same argument applied to (\ref{0n01}) entails
that the group $e^{itM_{12}}$, $t \in \RR$, is contained in $\Us_{\Ns}$.
One may then apply the same reasoning to the equality
\begin{equation}
e^{isM_{12}}e^{itM_{01}}e^{-isM_{12}} = e^{it(\cosh(s) M_{01} - \sinh(s) 
M_{02})}\end{equation}
to conclude that also the group $e^{itM_{02}}$, $t \in \RR$, is contained 
in $\Us_{\Ns}$. The assertion now follows for $n = 2$.
\end{proof}

\begin{lemma} \label{invariance}
Let $\Psi \in \Hs$ satisfy $U(\lambda_{01}(t))\Psi = \Psi$,
for all $t \in \RR$. Then $U(\lambda)\Psi = \Psi$, for all 
$\lambda \in \AdSG$. 
\end{lemma}

\begin{proof} Consider first $n \geq 3$. Using the equation
\begin{equation} \label{010j}
e^{isM_{01}}e^{itM_{0j}}e^{-isM_{01}} = 
e^{it(\cosh(s) M_{0j} + \sinh(s) M_{1j})}
\, , \quad j = 2,\ldots,n-1 \, ,
\end{equation}
and setting $t = 2re^{-\vert s \vert}$, the continuity 
of the representation $U(\AdSG)$ implies that
\begin{equation} \label{equation10}
\lim_{s \rightarrow \pm\infty} e^{isM_{01}}e^{i2re^{-\vert s \vert}M_{0j}}e^{-isM_{01}} = e^{ir(M_{0j} \pm M_{1j})} \, ,
\end{equation}
for $j = 2,\ldots,n-1$, where the limit is the strong operator limit on $\Hs$.
But then, by hypothesis, one has
\begin{equation}
\lim_{s \rightarrow \pm\infty} \Vert e^{isM_{01}}e^{i2re^{-\vert s \vert}M_{0j}}e^{-isM_{01}}\Psi - \Psi\Vert = 
\lim_{s \rightarrow \pm\infty} \Vert e^{i2re^{-\vert s \vert}M_{0j}} \Psi - \Psi \Vert = 0 \, ,
\end{equation}
since $e^{i2re^{-\vert s \vert}M_{0j}}$ converges strongly to the identity 
$1$ on $\Hs$ as $s \rightarrow \pm\infty$. These two equations then imply
\begin{equation}
e^{ir(M_{0j} \pm M_{1j})}\Psi = \Psi \, ,
\end{equation}
for all $r \in \RR$ and $j = 2,\ldots,n-1$. Similarly, the equation
\begin{equation} \label{011n}
e^{isM_{01}}e^{itM_{1n}}e^{-isM_{01}} = e^{it(\cosh(s) M_{1n} + \sinh(s) M_{0n})}
\end{equation}
yields
\begin{equation}
e^{ir(M_{1n} \pm M_{0n})}\Psi = \Psi \, ,
\end{equation}
for all $r \in \RR$. By using the Trotter product formula again and taking
suitable limits, it is clear that $M_{0j}\Psi = 0 = M_{1j}\Psi$, for all 
$j = 0,\ldots,n$. Equation (\ref{liealgebra}) then implies that $\Psi$ 
is annihilated by all of the generators $M_{\mu\nu}$, which yields the 
assertion for $n \geq 3$.

     If $n = 2$, equation (\ref{011n}) yields 
\begin{equation}
e^{ir(M_{12} \pm M_{02})}\Psi = \Psi \, ,
\end{equation}
for all $r \in \RR$, and thus $M_{12}\Psi = 0 = M_{02}\Psi$. The assertion
then follows for $n = 2$.
\end{proof}

\noindent We mention that Lemma \ref{invariance} is also stated
(without explicit proof) as Lemma 5.4 in \cite{BEM}.

\section{Wedge inclusions in AdS} \label{sectioninclusion}
\setcounter{equation}{0} 

     Here we give the proof of some useful geometric properties of the
subregions of \AdS which we have identified as the correct choice of
wedges in \AdS. Indeed, we wish to show that for any wedge $W \in \Ws$
there exist wedges $W_0 \in \Ws$ 
which are properly contained in $W$, $W_0 \Subset W$, respectively,
such that ${W_0}^\prime$ is
properly spacelike separated from $W$,
$W \Subset W_0$. In fact, such wedges are quite
abundant.  These results are to be contrasted with the situation in de
Sitter space, where de Sitter wedges satisfy $W_1 \subset W_2$ if and
only if $W_1 = W_2$ \cite{BDFS}. In Minkowski space of
dimension $n \geq 3$, two wedges form
an inclusion $W_1 \subset W_2$ only if $W_1$ is a suitable translation
of $W_2$. Hence, also in the latter case there do not exist properly
spacelike separated wedges. 

     It is convenient to use the following description of $W_R$:
\begin{equation}
W_R = \{ x \in \AdS \mid e_{\pm} \cdot x < 0 \, , \, x \cdot e_4  > 0 \} \, , 
\end{equation}
where $e_\pm = (\pm 1,1,0, \dots ,0)$ and $e_4 = (0, \dots ,0,1)$. Consider
the lightlike vectors $f_\pm = (\pm 1, c, 0, \dots ,0, s)$, where $s > 0$ and 
$c^2 = 1 + s^2$, and the wedge they determine:
\begin{equation}
W_0 \doteq \{ x \in \AdS \mid x \cdot f_\pm < 0\, , x \cdot e_4 > 0 \} \, .
\end{equation}
The edge of this wedge is the spacelike submanifold 
\begin{equation}
\Es_1 \doteq \left\{ 
(0,s(1+\sigmavec^2)^{1/2}, \sigmavec ,c(1+\sigmavec^2)^{1/2}) 
\mid \sigmavec \in \RR^{n-2} \right\} \, ,
\end{equation}
which is contained in $W_R$. 

\begin{lemma} \label{neighborhood}
With the above definitions, for any $t \in \RR$ there exists a
neighborhood $\Ns$ of the identity in $\AdSG$ such that
$\lambda \lambda_{01}(t)W_0 \subset W_R$
for all $\lambda \in \Ns$. Hence,
for any wedge $W_2 \in \Ws$ there exist wedges $W_1,W_3 \in \Ws$ such that
$W_1 \Subset W_2$ and $W_2$ and $W_3$ are properly spacelike separated.
\end{lemma}

\begin{proof} In order to show that 
$\lambda \lambda_{01}(t)\overline{W_0} \subset W_R$ for
$\lambda$ in a neighborhood of the identity, it suffices to show that
the characteristic boundary of $\lambda \lambda_{01}(t)\overline{W_0}$ 
is contained in $W_R$: 
$\lambda \lambda_{01}(t)(\Es_1 + \RR_+ f_\pm) \subset W_R$, {\it i.e.}
$\lambda\lambda_{01}(t)(x + l f_\pm) \in W_R$, for all $l \geq 0$, 
$x \in \Es_1$.
Since $\lambda$ is to be in a neighborhood of the identity $1$, consider
$\lambda = 1 + M$, where $\Vert M \Vert < \varepsilon$ and $\Vert\cdot\Vert$
is the norm on the $(n+1) \times (n+1)$ matrices with real entries. 
Then one has
\begin{equation}
\begin{split}
& \lambda\lambda_{01}(t)(x + l f_+) \cdot e_\pm =  
(\pm \sinh(t)  -  \cosh(t)) s(1 +\sigmavec^2)^{1/2} \\
& + l(\pm \cosh(t) \pm c \sinh(t) - \sinh(t) -c \cosh(t)) +  
  M \lambda_{01}(t) x \cdot e_\pm +
   lM \lambda_{01}(t) f_+ \cdot e_\pm \label{hier} \\
& \leq (\pm \sinh(t)  - \cosh(t)) 
s(1+\sigmavec^2)^{1/2} +  
\Vert M \Vert \, \Vert e_\pm \Vert \{ \Vert \lambda_{01}(t) x \Vert + l 
\Vert \lambda_{01}(t) f_+ \Vert\} \, ,
\end{split}
\end{equation}
where use was made of the fact that 
\begin{equation}
\pm \cosh(t) \pm c \sinh(t) - \sinh(t) -c \cosh(t) =
(\mp 1 + c)(\pm \sinh(t) - \cosh(t)) < 0 \, .
\end{equation}
But 
$\Vert \lambda_{01}(t) f_\pm \Vert \leq  2\cosh(t)$, 
$\Vert \lambda_{01}(t) x \Vert \leq \cosh(t)(1 + c^2 + 2s^2)^{1/2}
(1+\sigmavec^2)^{1/2}$ and
$\Vert e_\pm \Vert = \sqrt{2}$.
In addition, $\pm \sinh(t) - \cosh(t) < 0$, for all $t \in \RR_+$.
Hence, if $\varepsilon$ is sufficiently small, there exists a 
$\delta < 0$ (depending on $t$) such that
\begin{equation}
\lambda \lambda_{01}(t)(x + l f_+) \cdot e_\pm \leq \delta < 0 \, , \quad
x \in \Es_1 \, , l \geq 0 \, .
\end{equation}
Similarly, one shows that
\begin{equation}
\lambda \lambda_{01}(t)(x + l f_-) \cdot e_\pm \leq \delta < 0 \, , \quad
x \in \Es_1 , l \geq 0 \, ,
\end{equation}
for suitably small $\varepsilon$.

     Since $(\lambda W)^\prime = \lambda W^\prime$, for all 
$\lambda \in \AdSG$ 
and $W \in \Ws$, it is clear that $W_2$ and $W_3^\prime$ are properly
spacelike separated if and only if $W_2 \Subset W_3$. Thus,
since $\Ws = \{ \lambda W_R \mid \lambda \in \AdSG \}$,
the remaining assertions follow at once.
\end{proof}

     It is of interest to note that in $\AdS$, $n \geq 3$, 
there exists a wedge $W_2 \in \Ws$ such that $\overline{W_2} \subset W_R$, 
but in any neighborhood of the identity of $\AdSG$ there exists some
$\lambda$ such that $\lambda W_2 \not\subset W_R$. 

\section{The Reeh--Schlieder property} \label{sectionReehSchlieder}
\setcounter{equation}{0} 

     We prove that the theories we are considering here must satisfy
the Reeh--Schlieder property for the wedge algebras. 
Let $W \in \Ws$ be a wedge, and let $\Bs(W)$ denote the 
*--algebra consisting of all $B \in \As(W)$ for which there exists a
neighborhood $\Ns(B)$ of the identity in $\AdSG$
such that $B(\lambda) \doteq U(\lambda)BU(\lambda)^{-1} \in \As(W)$, for
all $\lambda \in \Ns(B)$. Note that Lemma \ref{neighborhood} entails
that there exists a wedge $W_0 \in \Ws$ such that $\As(W_0) \subset \Bs(W)$. 

\begin{prop} \label{ReehSchlieder}
Let Assumptions (i)--(iv) obtain. Then 
$\Omega$ is cyclic for $\As(W)$, given any $W \in \Ws$.
\end{prop}

\begin{proof}\footnote{This proof is a straightforward adaptation of an 
argument given in \cite{BB}.} 
Let $W,W_0 \in \Ws$ and $\Bs(W)$ be as described above and let $\Ns$
be a neighborhood of the identity in $\AdSG$ such that 
$\lambda^{-1} W_0 \subset W$, for all $\lambda \in \Ns$. By the 
covariance assumption (iii), it suffices to consider $W = W_R$. 
Further, let $\Psi\in\Hs$ be orthogonal to the set of vectors 
$\As(W)\Omega$. 

     Since $\Bs(W) \subset \As(W)$, $\Psi$ is also orthogonal to 
$\Bs(W)\Omega$. {}From the definition of $\Bs(W)$ and the continuity of
the representation $U$, it is clear that for any $B \in \Bs(W)$ and
$\lambda \in \Ns$ as described above there exists an $\varepsilon > 0$ 
such that 
$B(\lambda\lambda_{01}(t)\lambda^{-1}) \in \Bs(W)$, for all
$\vert t \vert < \varepsilon$. Therefore, one has
\begin{equation}
\langle\Psi,B(\lambda\lambda_{01}(t)\lambda^{-1})\Omega\rangle = 0 \, ,
\end{equation}
for $\vert t \vert < \varepsilon$. Since 
$\lambda\lambda_{01}(t)\lambda^{-1}W_0 \subset \lambda W$, one also has
$B(\lambda\lambda_{01}(t)\lambda^{-1}) \in \As(\lambda W)$, for all
$t \in \RR$. Hence, by the KMS property of the restriction of $\omega$ to
$\As(\lambda W)$, the function
\begin{equation}
t \mapsto B(\lambda\lambda_{01}(t)\lambda^{-1})\Omega \, , \quad t 
\in \RR \, ,
\end{equation}
extends analytically to a vector--valued function in the strip
$\{ z \in \CC \mid 0 < \Im(z) < \beta/2 \}$ with continuous boundary values.
Therefore, one must have
\begin{equation}
\langle U(\lambda\lambda_{01}(t)\lambda^{-1})^{-1} \Psi, B\Omega\rangle
= \langle\Psi,B(\lambda\lambda_{01}(t)\lambda^{-1})\Omega\rangle = 0 \, ,
\end{equation}
for all $t \in \RR$ and $B \in \Bs(W)$. By iterating this argument, it
follows that for $\lambda_1,\ldots,\lambda_k \in \Ns$ and
$t_1,\ldots,t_k \in \RR$, 
\begin{equation}
\langle U(\lambda_1\lambda_{01}(t_1)\lambda_1^{-1})^{-1}\cdots
U(\lambda_k\lambda_{01}(t_k)\lambda_k^{-1})^{-1} \Psi, B\Omega\rangle
= 0 \, .
\end{equation}
Lemma \ref{generate} then implies that $U(\lambda)\Psi$ is orthogonal to
$\Bs(W)\Omega$, for any $\lambda \in \AdSG$, and hence $\Psi$ is orthogonal
to $U(\lambda)^{-1}\Bs(W)\Omega = \Bs(\lambda^{-1}W)\Omega$.

     Moreover, since $\Bs(W)$ is a *--algebra, $B^* U(\lambda)\Psi$
is orthogonal to $\Bs(W)\Omega$, for any $B \in \Bs(W)$ and
$\lambda \in \AdSG$. Hence, by induction, for any
$\lambda_1,\ldots,\lambda_k \in \AdSG$ and $B_1,\ldots,B_k \in \Bs(W)$
one has 
\begin{equation}
\langle\Psi,B_1(\lambda_1)\cdots B_k(\lambda_k)\Omega\rangle = 0 \, .
\end{equation}
Putting these results together, it now follows that $\Psi$ is orthogonal
to \newline
$\big( \bigcup_{\lambda\in\text{SO}_0(2,n-1)} U(\lambda)\Bs(W)U(\lambda)^{-1} \big)
\Omega$. And since $\As(W_0) \subset \Bs(W)$, one observes 
\begin{equation}
\big( \bigvee_{\lambda\in\text{SO}_0(2,n-1)} 
\Bs(\lambda W)\big)\Omega \quad \supset 
\big( \bigvee_{\lambda\in\text{SO}_0(2,n-1)} 
\As(\lambda W_0)\big)\Omega = \As\Omega 
\, ,
\end{equation}
by the assumption of weak additivity (\ref{weakadditivity}). Thus, 
$\Psi$ is orthogonal to $\As\Omega$. As $\Omega$ is cyclic for
$\As$, the assertion is proven.
\end{proof}

\section{Multiplicity of energy levels and nuclearity} \label{multiplicity}
\setcounter{equation}{0} 

     In this appendix we wish to prove that given any  
irreducible unitary positive energy representation $U_1$ of the  
anti--de Sitter group $\AdSG$ on a Hilbert space $\Hs_1$, then the
corresponding unitary representation $U$ obtained by 
``second quantization'' on the bosonic or fermionic Fock 
space $\Hs$ based upon $\Hs_1$ satisfies condition (NC). Hence
the free field examples discussed in the main text and 
Appendix \ref{examples} satisfy (NC). To verify this, we shall 
have need of some basic results about unitary representations 
of $\AdSG$. We merely recall these and refer the reader to
\cite{F1,F2,Ev} for details and proofs. We shall present the case 
$n = 4$; the situation is similar for the other cases.

     Let $SO(2)\otimes SO(3) \subset \AdSGfour$ be the maximal compact 
subgroup consisting of the ``time rotations'' around 
AdS$^4$ and the spatial rotations. Restricting $U_1$ to this
compact group, the Hilbert space $\Hs_1$ decomposes into
a direct sum of corresponding irreducible subspaces which are labelled by
the energy $n \in \IN$ and angular momentum $l \in \IN_0$.
It has been shown in \cite{Ev} that in this decomposition
there appear only representations
where $l < n$. Moreover, for given $(n,l)$, the number of these
representations is bounded by $l+1$. Taking these facts into
account, the multiplicity $\mu_n$ of the eigenvalue $n$
of the generator of the time rotations on $\Hs_1$ can be estimated
by $\mu_n \leq \sum_{l=0}^{n-1} \, (2l +1)(l+1) \leq 2 n^3$,
$n \in \IN$.

Next, let $M_{0 4}$ be the generator of the time rotations on the 
Fock space $\Hs$. The bound given above entails by standard
arguments in statistical mechanics that in the bosonic case 
the corresponding partition function satisfies, 
for any $\gamma > 0$,
\begin{equation} \label{nucl1}
\begin{split}
\ln {\rm tr} e^{-\gamma M_{04}} 
& = - \sum_n \mu_n  \,\ln(1 - e^{-\gamma n}) 
\leq 2 \sum_n n^3  \, \frac{e^{-\gamma n}}{1 - e^{-\gamma}} \\
& = \frac{2}{1 - e^{-\gamma}} (- \, \partial_\gamma)^3
\frac{1}{1 - e^{-\gamma}} \leq \frac{12 e^{-\gamma}}{(1 - e^{-\gamma})^5}
\leq \frac{12 \cdot 5^5}{\gamma^5} \, .
\end{split} 
\end{equation}
Denoting by $\boldsymbol{\mu}_n$ the multiplicity of the eigenvalue
$n$ of $M_{0 4}$, $n\in \IN_0$, we obtain the estimate 
\begin{equation}
{\rm tr} \, e^{-\gamma M_{04}} = \sum_{n} \boldsymbol{\mu}_n 
e^{- \gamma  n} \leq e^{12 \cdot 5^5 / \gamma^5} \, ,
\end{equation}
which implies 
\begin{equation}
\boldsymbol{\mu}_n  \leq e^{\gamma n + 12 \cdot 5^5 / \gamma^5} \, ,
\end{equation}
for all $n \in \IN_0$ and $\gamma > 0$. With the choice
$\gamma = 5 n^{-1/6}$, we conclude that 
\begin{equation}
\boldsymbol{\mu}_n \leq e^{17 \,  n^{5/6}} \, , \quad n \in \IN_0 \, .
\end{equation}
A similar argument applies also to the fermionic case
and in any number of spacetime dimensions. Hence we have the following result.

\begin{prop}
In any free boson or fermion model based upon an irreducible positive
energy representation of the anti--de Sitter group, condition (NC)
holds.
\end{prop}

\section{Examples} \label{examples}
\setcounter{equation}{0}

     In this appendix we shall discuss some examples of nets and
states which fulfill Standing Assumptions (i)--(iv), as well as the
condition (NC). Because AdS is not globally
hyperbolic, the standard means of obtaining examples do not
suffice. Free field models on AdS have been discussed in a series of
papers by Fronsdal \cite{F1,F2,F3,F4} and by Avis, Isham and Storey
\cite{AIS}. More recently, the Wightman functions of quantum field
models on AdS satisfying certain general conditions have been treated
rigorously in \cite{BEM}. In addition, models of quantum field
theories on AdS can also be obtained {\it via} holography
\cite{Rehren} (see also \cite{BBMS}).  We begin our discussion with
the latter.

     In an elegant paper \cite{Rehren} Rehren has given rigorous
mathematical meaning to the notion of holography, namely the
correspondence between theories on $\AdS$ and conformally invariant
theories on the boundary $CM_{n-1}$ of \AdS, compactified
$(n-1)$--dimensional Minkowski space. He shows that between the set of
wedges $W$ in AdS and the set of (conformal images of) double cones
$C$ in the boundary there exists a canonical bijection $\alpha$ which
preserves inclusions and causal complements, and intertwines the
actions of the anti-de Sitter group and of the conformal group (which
are isomorphic groups):
\begin{equation}
\alpha(\lambda(W)) = \tilde{\lambda}(\alpha(W)) \, , \quad
\alpha^{-1}(\tilde{\lambda}(C)) = \lambda(\alpha^{-1}(C)) \, ,
\end{equation}
where $\tilde{\lambda}$ is the restriction of the action of 
$\lambda \in \AdSG$ to the boundary. The double cone $C = \alpha(W)$ is 
defined to be the intersection of $W$ with the boundary. Hence, given a net 
$\{ \widetilde{\As}(\alpha(W)) \}$ 
associated with a (for example) free quantum field 
on \textit{CM}$_{n-1}$ and a vacuum 
state $\tilde{\omega}$ on this net, one can 
define a net $\{ \As(W)\}$ and state $\omega$ on $\AdS$ by
\begin{equation}
\As(W) \doteq \widetilde{\As}(\alpha(W)) \, , \quad 
\omega(A) \doteq \tilde{\omega}(A) \, , 
\end{equation} 
for every $A \in \As(W) = \widetilde{\As}(\alpha(W))$. 
{}From the results in \cite{Rehren}
it is easy to show that the resulting theory on \AdS fulfills our 
assumptions (i)--(iv), provided the underlying net on 
$CM_{n-1}$ complies with the standard assumptions of 
conformal Minkowski space theories. 
In such theories one also has the equality
$\As(W) = \As(-W)$, for every $W \in \Ws$. The CGMA and the \msc, 
formulated in \cite{BDFS}, both obtain in these models.

     Using an irreducible representation of $\AdSGfour$,
Fronsdal \cite{F2} defines Hermitian free fields on the
covering space of AdS$^4$; only if the energy spectrum 
of the theory is contained in $\IN_0$ does his field restrict
to AdS proper. In this latter case, Fronsdal's model satisfies
assumptions (i)-(iv), as well as (NC) -- see below and Appendix
\ref{multiplicity}. Moreover, the elements of $\As(W)$ commute 
weakly with those of $\As(\widetilde{W}')$. Yet if the energy spectrum 
is not a subset of $\IN_0$, this feature is no longer present, confirming 
our expectation that Theorem \ref{weaklocality2} can not hold in general
for fields not manifesting the periodicity in time required to enable
them to be defined on AdS. We note that since in his examples the 
fields are invariant under the map $x \mapsto -x$, one has 
$\As(\widetilde{W}') = \As(W') = \As(W)'$.

     In \cite{AIS} Avis, Isham and Storey use an embedding of (the
covering space of) AdS into the static Einstein universe to construct
a free quantum field on AdS. Since the static Einstein
universe is globally hyperbolic, one can rigorously construct free
fields and the associated nets of local algebras in (subsets of) that
spacetime. And since the covering space of AdS can be conformally
embedded into the Einstein universe, one can define free massless, \ie
conformally invariant, fields on AdS. However, the matter is
complicated by the fact that one must find suitable boundary
conditions at spacelike infinity in AdS -- we must refer the reader
to \cite{AIS} for details. Since the resultant fields manifest the
necessary periodicity in the time variable, they may be understood to
be defined on AdS.  Inspection of the resulting representations then
led them to construct analogous representations of ``conformally
coupled massive'' free fields on AdS. {}From the construction of their
examples and the results in \cite{BEM} it follows that Standing 
Assumptions (i)--(iv) hold. All of their examples are 
consistent with Theorem \ref{weaklocality2}.

     In \cite{BEM} an axiomatic study is made of quantum fields on the
covering space of AdS from the point of view of a suitable
modification of the Wightman function approach to quantum field theory
on Minkowski space. Explicit examples of two-point functions
satisfying their assumptions are given there, which include Fronsdal's
examples. A subclass of those two-point functions restrict to AdS; it
is of interest to note that such two-point functions are characterized
by a certain uniformity property --   cf. Section 6 in
\cite{BEM}. Using the results of \cite{PuWo,BEM} it is easy to show
that free fields built upon those two-point functions satisfy our
assumptions (i)-(iv); the elements of a subclass also satisfy
assumption (NC). The authors of \cite{BEM} also observe that the
locality property proven in Theorem \ref{weaklocality2} obtains only
in those free field models which restrict to AdS proper \cite[p.\ 509]{BEM}.

\end{document}